\documentclass[reprint]{revtex4-1}
\usepackage{fullpage}
\usepackage{amsfonts}
\usepackage{array}
\usepackage{natbib}		
\usepackage{graphicx}	
\usepackage{subcaption}
\usepackage{caption}
\usepackage{amsmath}
\usepackage{xcolor}
\usepackage{multirow}
\usepackage{lipsum}% http://ctan.org/pkg/lipsum

\begin{document}

\title{Capillary-bridge Forces Between Solid Particles: Insights from Lattice Boltzmann Simulations}
\author{Lei Yang}
\author{Marcello Sega}
\affiliation{Helmholtz Institute Erlangen-Nürnberg for Renewable Energy (IEK-11), Forschungszentrum Jülich, Fürther Stra{\ss}e 248, 90429 Nürnberg, Germany}
\author{Jens Harting}
\affiliation{Helmholtz Institute Erlangen-Nürnberg for Renewable Energy (IEK-11), Forschungszentrum Jülich, Fürther Stra{\ss}e 248, 90429 Nürnberg, Germany}
\affiliation{Department of Chemical and Biological Engineering and Department of Physics, Friedrich-Alexander-Universität Erlangen-Nürnberg, Fürther Stra{\ss}e 248, 90429 Nürnberg, Germany} 

\begin{abstract}
Liquid capillary-bridge formation between solid particles has a critical influence on the rheological properties of granular materials and, in particular, on the efficiency of fluidized bed reactors. The available analytical and semi-analytical methods have inherent limitations, and often do not cover important aspects, like the presence of non-axisymmetric bridges. Here, we conduct numerical simulations of the capillary bridge formation
between equally and unequally-sized solid particles using the lattice Boltzmann
method, and provide an assessment of the accuracy of different families of analytical models. We find that some of the  models taken into account are shown to perform better than others. However, all of them 
fail to predict the capillary force for contact angles
larger than $\pi/2$, where a repulsive capillary force attempts to push the solid particle outwards to minimize the surface energy, especially at a small separation distance. We then apply the most suitable model to study the impact of capillary interactions on particle clustering using a coupled lattice Boltzmann and Discrete Element method.

% Please include a maximum of seven keywords
\end{abstract}
\maketitle

\section{Introduction}

In many industrial processes such as wet granulation, coating, or
drying, powders can contain a small amount of liquid. In these powders, the appearance of liquid capillary bridges between the grains generates adhesion forces at the micro-scale that  
can modify dramatically the granular medium's mechanical properties
at the macro-scale~\cite{herminghaus2005}. One of the signatures of capillary bridges
is the formation of agglomerates. Next to many other applications, this is for example important in the combustion of biomasses. The ash composition of biogenic fuels like energy crops or agricultural residues contains always a high amount of silicon and potassium, which are prone to melt at the
operating temperatures of fluidized beds~\cite{gatternig2015investigations}. The molten ash then forms
liquid bridges between particles and causes agglomeration. The growth of agglomerates, if not counteracted properly, can cause the defluidization of the bed. In this
sense, understanding the interaction between solid particles as
mediated by the liquid bridges is critical to reduce the
limiting factors of fluidized bed reactors and to improve their
performance. 

In the literature one can find a considerable amount of research focusing on analytical expressions that estimate the capillary forces induced by the bridges~\cite{lian1993theoretical,mikami1998numerical,willett2000capillary,
rabinovich2005capillary,megias2009capillary, megias2010analysis}.
Most of these works make use of either an energetic route (i.e., via the derivative of the total interface energy) or, equivalently\cite{lambert2008comparison}, approximate solutions of the Young-Laplace equation. Typically, once the meniscus geometry is known, the capillary force can be
obtained by multiplying the cross-sectional area by the Laplace
pressure and adding the surface tension of the liquid. 
The two main approximations employed to describe the meniscus shape are the toroidal~\cite{haines1925studies,fisher1926capillary} and the Derjaguin~\cite{derjaguin} approximations. Several other approximate solutions of the Young-Laplace equation based on these two have been reported in  literature~\cite{willett2000capillary,rabinovich2005capillary,butt2009normal}. 

Many of these approximations assume the bridge connecting two identical spherical particles. In reality, particles are often polydisperse in size, and some authors
proposed theoretical models for the prediction of the capillary
force for unequally-sized
spheres~\cite{willett2000capillary,chen2011liquid,sun2018liquid,lian2016capillary}. 
In addition, most studies found in literature
focus on capillary bridges with small, typically less than 10\%, liquid-to-solid volume ratios  (e.g.,\cite{huppmann1975modelling, lian1993theoretical,willett2000capillary,rabinovich2005capillary, roy2016micro}), whereas  only few studies focus  on larger volumes~\cite{sun2018liquid,farmer2015asymmetric,xiao2020capillary}. Interestingly, with a large enough liquid volume, the solution that minimizes the surface energy is not anymore axisymmetric~\cite{vogel2006convex}, as it was also shown numerically and experimentally~\cite{farmer2015asymmetric}. Moreover, when increasing the separation distance, it is possible to observe a transition from a convex to a concave profile~\cite{xiao2020capillary}. 

Numerical methods to tackle the problem of the liquid capillary bridge were devised already in the early 1970s by Erle and coworkers, obtaining a satisfactory agreement at large separation distances~\cite{erle1971liquid}. Generally, for static solutions, one can numerically minimize the interfacial free energy using a number of freely available tools~\cite{brakke1992surface}. If the dynamics of the bridge formation does play a role, CFD methods such as the volume-of-fluid (VOF) and level set methods are common choices~\cite{rider1998reconstructing,sussman1994level}. They solve the macroscopic Navier-Stokes equations by tracking or capturing the interface. An alternative approach is the lattice Boltzmann method (LBM), which became popular due to its straightforward implementation, its inherent parallelization and a wide selection of multiphase models being available~\cite{kruger2017lattice,liu2016multiphase}. 
We would like to emphasize that the LBM can be used to model
dynamic problems that are beyond of the scope of other methods based just on the minimization of surface energy as, for example, implemented in Surface Evolver~\cite{brakke1994surface}. In the LBM, hydrodynamics is being described at the Navier-Stokes level in the nearly incompressible limit. The dynamics of the solid particles can be simulated by solving Newton’s equations of motion for the translational and rotational degrees of freedom, and coupled to the single or multicomponent fluid using various approaches\cite{ladd1994numerical,ahlrichs1999simulation,sega2013mesoscale}.  
In comparison with conventional CFD methods, the LBM is limited to low Mach numbers. Also, at high  Knudsen numbers, more work is needed to improve the accuracy of the method while maintaining computational efficiency. However, these limitations have no influence on the current study of capillary bridges.

%Up to now, the current multiphase/multicomponent lattice models are preferly suitable for isothermal problems. The development of a reliable LBM for thermal systems in the application of reactive flows would be challenging ~\cite{aidun2010lattice}. 

In this work, we use the multicomponent inter-particle
potential model by Shan and Chen~\cite{shan1993lattice} coupled to solid particles modelled using the approach introduced by Ladd and Aidun~\cite{ladd1994numerical, ladd2001lattice,aidun1998direct,HFRRWL14} to investigate the validity and suitability of several analytical approximations of the bridge forces.

This paper is organized as follows. In Section II(A-B), we introduce some widely used models for the capillary bridge force. In section II(C), we describe the multicomponent lattice Boltzmann method that we use to model the capillary bridges forming between spherical particles. In Section III we investigate five different capillary bridge models with small and large bridge volumes, some of which have been frequently used to describe capillary bridge forces in Discrete Element Method (DEM) simulations~\cite{roy2016micro,roy2017}. As an example of the possible applications, we use one of the potentials to model the interaction between particles and show the formation of clusters due to the capillary bridges under shear flow. In Section IV we discuss the limit of the theoretical models and summarize our results.
\section{Methodology}
\subsection{Capillary bridge geometry and bridge volume}

\begin{figure}
\includegraphics[width=\columnwidth]{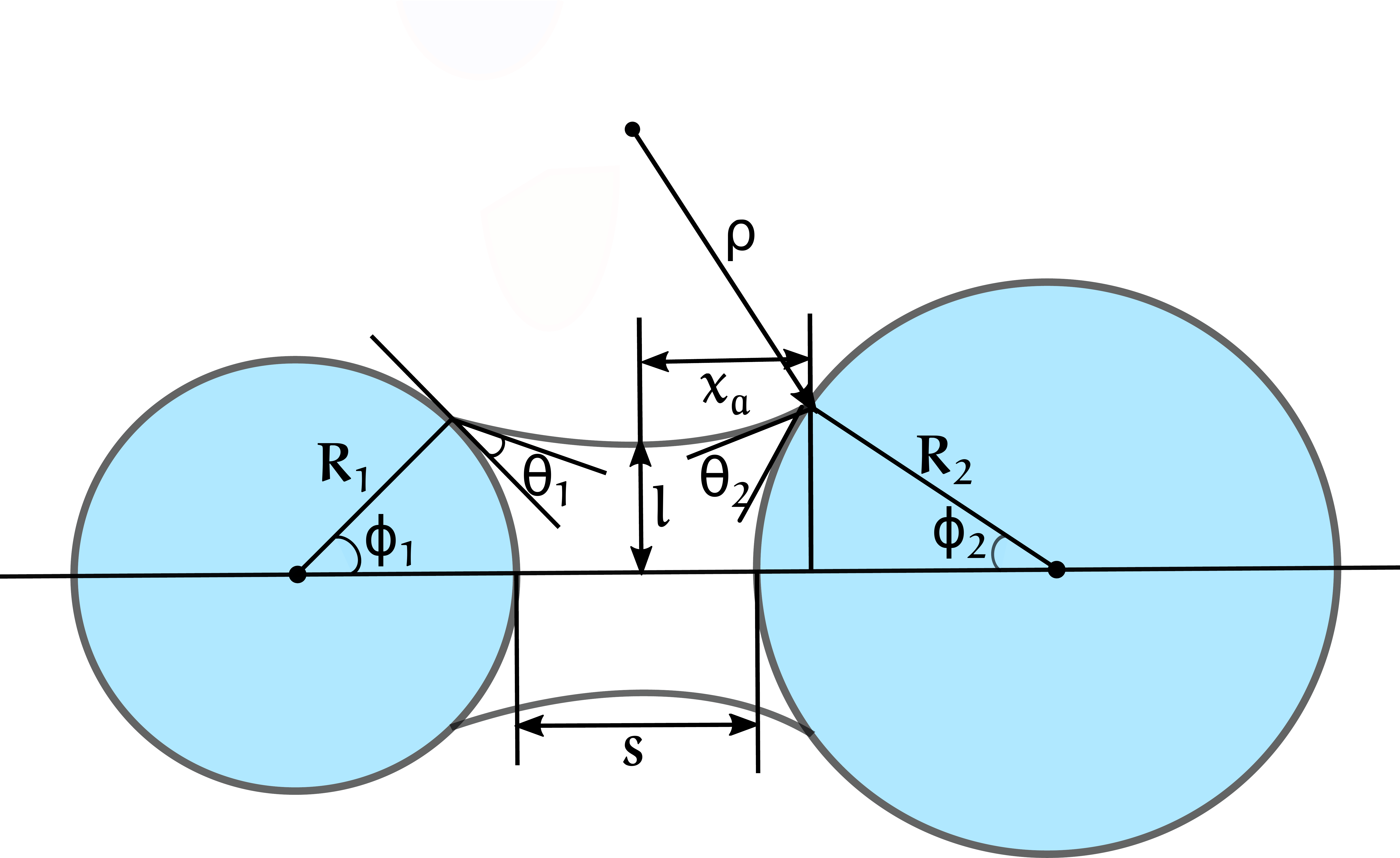}
\caption{Sketch of the liquid bridge geometry displaying the parameters involved in the calculations.}
\label{fig:sketch}
\end{figure}

The volume of the liquid bridge is an important ingredient in the calculation of the capillary force and needs special attention. 
Fig.~\ref{fig:sketch} depicts a sketch of the main geometrical quantities involved in the description of the bridge. Two particles of radius $R_1$
and $R_2$, respectively, are separated by the surface-to-surface distance $S$. The principal radii of the liquid meniscus are $\rho$
and $l$, respectively, and the volume of each particle is $V_{p,1}$ and $V_{p,2}$. The contact angles
are denoted as $\theta_1$ and $\theta_2$, and the half-filling angles are
denoted as $\phi_1$ and $\phi_2$. $x_a$ and $y_a$ are the abscissa and ordinate of the contact point between the solid and liquid profiles. The coordinate system is chosen such that the $x$-axis lies along the straight
line joining the centers of the particles, with its origin located at the middle point between the spherical surface. 
By integrating the liquid profile $y(x)$ one can obtain the liquid volume of the bridge $V_l$, which in the case of two equal spheres reads
\begin{equation}\label{eq:1}
V_l= \pi \int_{-x_a}^{x_a} y^2(x) dx - 2V_s,
\end{equation}
where $V_s$ is the volume of the spherical caps wet by the liquid. The volume of the spherical caps are calculated geometrically as $\pi R_1^3 (2+\cos \phi_1)(1-\cos \phi_1)^2/3$ and $\pi R_2^3(2+\cos \phi_2) (1-\cos \phi_2)/3$, respectively. 
In the toroidal approximation, 
the meridional profile of pendular bridges is approximated by an arc of a circle.  This approximation leads to a
simple closed-form solution for the Young-Laplace equation, which has been widely used for the
estimation of the forces between spherical
particles~\cite{lian1993theoretical,huppmann1975modelling}. In this
approach, the surface curvature is not constant along the meniscus of the liquid bridge. Consequently, the total force in the toroidal approximation is a function of the local surface curvature and thus of the position of the orthogonal plane~\cite{willett2000capillary}. The toroidal approximation can be used for small menisci, where the effect of gravity can be neglected. Deviations from the toroidal approximation agree with the results from the forces calculated numerically with the exact bridge shape \cite{butt2009normal}.
Many authors used the toroidal approximation to calculate the liquid bridge volume for equally and unequally-sized spheres~\cite{megias2010analysis,sun2018liquid,megias2009capillary,huppmann1975modelling,pietsch1967haftkraft}. Pietsch and Rumpf
reported the expression of the bridge volume with a wide range of partial wetting, as shown
in Table \ref{table1}~\cite{pietsch1967haftkraft}. Other authors derived simpler expressions
with further simplifications. However, their expressions underestimate~\cite{huppmann1975modelling,simons1994analysis} or overestimate~\cite{rabinovich2005capillary} the bridge volume, as compared to the values obtained in Refs.~\cite{pietsch1967haftkraft,megias2009capillary}. Megias-Alguacil and Gauckler
 calculated the liquid bridge volume in
terms of the abscissa $x_a$, the contact angle and the separation
distance~\cite{megias2009capillary}. In case of two equal spheres, their expression is fully equivalent to the one from Pietsch and Rumpf if the relation
\begin{equation}
\phi=\arccos\left[S/(2R_p)+1-x_a/R_p\right]
\end{equation} is used~\cite{pietsch1967haftkraft}. Chen and coworkers developed an improved mechanical model
which is capable of analyzing the force and the volume of the liquid
bridge by assuming that the liquid bridge profile is circular in
shape between two unequally-sized spherical particles~\cite{chen2011liquid}. Later, Sun and
Sakai proposed a more general expression for the  
capillary bridge volume which coincides with the one of Chen and coworkers if the spheres have the same contact angle~\cite{sun2018liquid}.

\begin{table*}
\begin{center}
\setlength\arrayrulewidth{1.0pt}
    \begin{tabular}{llll}
    \hline
    Models & & &Capillary bridge volume $V_l$  \\ \hline
    Pietsch \& Rumpf~\cite{pietsch1967haftkraft}
&$\frac{V_l}{2\pi}$&= & $\rho(\rho^2 + b^2 ) \cos(\theta + \phi) - \frac{1}{3} \rho^3 \cos^3(\theta + \phi)$\\
&&&$- \frac{1}{24}d_p^3 (2 + \cos \phi) (1 - \cos \phi)^2$\\
&&&$- b \left[\rho^2 \cos(\theta + \phi) \sin(\theta + \phi) 
+\rho^2 \left( \frac{\pi}{2} - \theta - \phi \right)\right]$\\
&$b$&=&$\rho + l$, $\rho = \frac{d_p (1 - \cos \phi) + S}{2 \cos(\theta + \phi)}$\\
&$l$&=&$R_p \sin \phi + \rho [\sin(\theta + \phi) - 1] $\\
Alguacil \& Gauckler~\cite{megias2009capillary}
 &$\frac{V_l}{2\pi}$&=&$((l-\rho)^2 + \rho^2)x_a -\frac{x_a^3}{3} + (l-\rho)\left[x_a \sqrt{\rho^2-x_a^2} + \rho^2 \arcsin \frac{x_a}{\rho} \right]$\\
&&&$-\frac{1}{3}(x_a-S/2)^2(3R_p -(x_a-S/2))$\\
&$\rho $&=&$ \frac{x_a R_p}{(S/2 + R_p -x_a)\cos \theta - \sqrt{R_p^2 - (x_a-S/2 - R_p)^2} \sin\theta}$\\
&$l $&=&$ \sqrt{R_p^2 - (x_a-S/2 - R_p)^2} + x \frac{\sqrt{R_p^2 - (x_a-S/2 - R_p)^2} \cos\theta + (S/2+R_p-x_a) \sin\theta -R_p}{(S/2+R_p-x_a) \cos\theta - \sqrt{R_p^2 - (x_a-S/2 - R_p)^2} \sin\theta}$\\
Chen et al.\cite{chen2011liquid}
&$V_1$&=& $\pi\rho(a^2+\rho^2)(\cos(\phi_1+\theta)+\cos(\phi_2+\theta))- \frac{\pi}{3}\rho^3(\cos^3(\phi_1 + \theta) $\\
&&&$+ \cos^3(\phi_2+\theta)) -a\pi \rho^2(\sin(\phi_1 + \theta) \cos(\phi_1 + \theta) + \sin(\phi_2 + \theta) \cos(\phi_2 + \theta)$ \\
&&&$+ a\rho^2( \phi_1 + \phi_2 +2\theta - \pi)) $\\
&$V_2$&=&$ \frac{\pi}{3}\left((2-3\cos\phi_1+\cos^3\phi_1)R_1^3 + (2-3\cos\phi_2+\cos^3\phi_2)R_2^3 \right) $ \\
&$V_l$&=&$V_1 - V_2$\\
&$\rho$&=&$\frac{R_1(1-\cos\phi_1)+ S+ R_2(1-\cos\phi_1)}{\cos(\phi_1 + \theta) + \cos(\phi_2 + \theta)}$\\
&$l$&=&$ R_1 \sin\phi_1 - \rho (1 -\sin(\phi_2 + \theta))$\\
%    %Sun et al.
%%&$\frac{V_l}{2\pi} = (\rho^3 + L \rho^2) (\cos(\theta + \phi) - (\frac{\pi}{2} - \theta - \phi)) + \frac{1}{2} \rho L^2 \cos(\theta + \phi)$\\
%%&$\rho = \frac{R_p(1- \cos \phi + S/2)}{\cos(\theta + \phi)}$\\
%%&$L = R_p \sin \phi - \rho (1 - \sin(\theta + \phi))$\\
    \hline
    \end{tabular}
\caption{Capillary bridge volume ($V_l$) from different theoretical models}
\label{table1}
\end{center}
\end{table*}

\subsection{Capillary bridge force}

The capillary force is usually calculated following one of two routes, namely,
 the geometric one (approximating the solution of the Laplace
equation) and the energetic one (taking the derivative of the total interfacial energy). In the geometric approach, the capillary force exerted on solid spheres due to the liquid bridge is the sum of two components, which are the contributions from the surface tension and the hydrostatic pressure, respectively. The surface tension and hydrostatic contributions can be obtained by knowing the height of the neck, the tangent of the profile and the contact area, respectively. The direction of the hydrostatic pressure force depends on the curvature of the meniscus, while the surface tension force is always attractive. Convex bridges yield positive Laplace pressure contributions and repulsive hydrostatic pressure forces. On the contrary, the pressure in concave bridges is negative, and the corresponding force attractive. Once the profile of the bridge is known, both terms can be determined by geometrical means. In this study, the capillary force is attractive for F > 0 and repulsive for F < 0.

In order to obtain the hydrostatic pressure in the geometric approach, the geometry of the bridge has to be approximated. The so-called toroidal and Derjaguin approximations are two widely applied options. For relatively small liquid bridges (mostly concave) and at stable separation distances, the toroidal approximation has been shown to produce errors smaller than 10$\%$ in the calculated capillary force, in comparison with exact numerical techniques \cite{lian1993theoretical,
hotta1974capillary,orr1975pendular}. However, with increasing
separation distance, the toroidal solution may underestimate the
capillary bridge force~\cite{mazzone1987behavior}. Alguacil and Gauckler studied the validity of the toroidal approximation for the case of convex liquid bridges, showing agreement within 30\% with the numerical solution of the Young-Laplace equation~\cite{megias2010analysis,megias2011accuracy}.

The Derjaguin approximation was originally developed for determining the force between unequal spheres based on the interaction energy between
planar surfaces. The validity of this approximation is intuitively limited to separation distances that are small compared to the radii of the spheres, or, equivalently, when the radius of the bridge profile is orders of magnitude smaller than the neck radius.
Rabinovich and coworkers found that Derjaguin's approximation is valid only at strictly zero separation distance~\cite{rabinovich2005capillary}. Willett and coworkers proposed a  variant of the Derjaguin approximation for the total capillary forces between spheres as a function of the separation distance and for a fixed bridge volume~\cite{willett2000capillary}. In the energetic approach, the total capillary force is obtained by perturbing the displacement $S$ between the two particles, provided
that the volume of the liquid is constant
\begin{equation}\label{eq:2}
F = \frac{dW}{dS},
\end{equation}
where $W$ is the interfacial free energy determined by surface tension, interface contact areas and contact angle \cite{israelachvili2015intermolecular}.  Rabinovich and coworkers showed that the energetic and geometric routes are equivalent despite the nonequilibrium nature of the problem, and proposed an explanation for the failure of the Derjaguin approximation at large distances~\cite{rabinovich2005capillary}.
Chau and coworkers calculated the capillary force for non-axisymmetrical shapes allowing the meniscus to fulfill the Kelvin equation~\cite{chau2007three}. More recently, Wang and coworkers used the interfacial energy minimization approach to study the forces and the rupture behaviour of water bridges between three spherical particles at equilibrium configurations~\cite{wang2017capillary}. 

In the following we consider five different theoretical models, which are representative of the bulk of works on the subject. Theoretical models have limitations due to the approximations made in solving the Young-Laplace equation. To gain insight into different models, we explore the quality of the models for the prediction of capillary bridges using our lattice Boltzmann simulations.

In model A, as reported by Lian and coworkers, the liquid bridge force is calculated by the toroidal approximation~\cite{lian1993theoretical}. We use the ``gorge'' method ($F=\pi l^2 \Delta p + 2\pi \sigma l $) from their work, in which the area at the neck is used to calculate the hydrostatic pressure force and the tangent at the neck to obtain the surface tension force. The capillary force is calculated in terms of the half-filling angle, separation distance, and contact angle. For a wide range of liquid bridge volumes and stable separation distances, this method produces errors within 10\% in the calculation of the capillary force in comparison withhttps://latex.hi-ern.de/project/5e565d1c2ef5060157deb605
those obtained from an exact numerical technique. 

Model B was used by several authors, deriving the expression for the capillary force by the toroidal approximation~\cite{megias2009capillary, huppmann1975modelling, pietsch1967haftkraft}. In this model, the surface tension force ($2\pi \sigma R_p \sin \phi \sin(\phi + \theta)$) is obtained at the intersection of the three-phase contact line; and the hydrostatic pressure force is calculated by the axially projected wetted area of the particle. In Model B, the capillary force depends on the half-filling angle, separation distance, and contact angle. 

Model C is a semi-analytical model proposed by Willett and coworkers~\cite{willett2000capillary}. The volume of the capillary bridge is in this case an input of the problem and, in the following comparison, we will use the value obtained from the simulations. The error of their approximations
is no more than 4$\%$ when the liquid-to-solid volume ratio is
0.001. However, the error increases with increasing volume ratio.
The accuracy can be improved with a more complex expression, which
is valid for half-filling angles $< 50^\circ$ and volume ratios $V^*$
less than 0.1, giving an error in the force estimation of less
than 3$\%$. Model C can also be used to calculate the total capillary force for unequal spheres. The authors pointed out that deviations from the solution for equal spheres occur only when the bridge volume is large compared to that of the spheres.

Model D represents the approach of Rabinovich and coworkers, who obtained the capillary force between two spheres based on the Derjaguin approximation~\cite{rabinovich2005capillary}. Model E is from Miakmi and coworkers, who derived a formula for the liquid bridge
force as an explicit function of the 
liquid bridge volume and separation distance based on the regression analysis of the numerical
solutions of the Young-Laplace equation~\cite{mikami1998numerical}. The details of the models
can be found in Table \ref{table2}. Again, like in the case of Model C, for both Model D and Model E, the liquid bridge volume is an input of the problem.

\begin{table*}
\setlength\arrayrulewidth{1.0pt}
\begin{center}
    \begin{tabular}{ll}
    \hline
    Models & Capillary force $F$  \\ 
    \hline
    Model A  
    &$F= 2 \pi \sigma  l \left(1 + \frac{l - \rho}{2 \rho} \right)$\\
 Lian et al.\cite{lian1993theoretical}  &$\rho = \frac{2R_p(1-\cos\phi)+S}{2\cos(\theta+\phi)}$\\
&$l = R_p \sin\phi - \left( R_p (1- \cos \phi) + S/2 \right) \frac{1- \sin(\theta+\phi)}{\cos(\theta+\phi)}$\\
    Model B   
    & $F =2 \pi R_p \sigma \sin\phi \sin(\theta + \phi) + \pi \sigma R_p^2 \sin^2\phi \left( \frac{1}{\rho} - \frac{1}{l} \right)$\\
Huppmann\cite{huppmann1975modelling}, Pietsch\cite{pietsch1967haftkraft}&$\rho = \frac{2R_p(1-\cos\phi)+S}{2\cos(\theta+\phi)}$\\
&$l = R_p \sin\phi - \left( R_p (1- \cos \phi) + S/2 \right) \frac{1- \sin(\theta+\phi)}{\cos(\theta+\phi)}$\\
 Model B    
    & $F = 2 \pi \sigma R_p\sin\phi \sin (\phi + \theta) - \pi R_p^2 \sigma \sin^2\phi \left( \frac{1}{l} + \frac{1}{\rho}\right)$\\
Megias \& Gauckler\cite{megias2010analysis}&$\rho = \frac{x_a R_p}{\sqrt{R_p^2 - (x_a-S/2 - R_p)^2} \sin\theta - (S/2 + R_p -x_a)\cos \theta }$\\
&$l = \sqrt{R_p^2 - (x_a-S/2 - R_p)^2} +$\\
& $\qquad{}+ x_a \frac{\sqrt{R_p^2 - (x_a-S/2 - R_p)^2} \cos\theta + (S/2+R_p-x_a) \sin\theta -R_p}{(S/2+R_p-x_a) \cos\theta - \sqrt{R_p^2 - (x_a-S/2 - R_p)^2} \sin\theta}$\\
    Model C  
    & $F = 2\pi R_p \sigma \exp\left[f_1 - f_2\exp\left(f_3 \ln\frac{S}{2\sqrt{V/R_p}} + f_4 \ln^2\frac{S}{2\sqrt{V/R_p}}\right)
    \right] $ \\ 
 Willett et al.\cite{willett2000capillary}&\\   
    Model D 
    & $F=\frac{2\pi \sigma R_p  \cos \theta}{1 + S/(2d)}$\\ 
Rabinovich et al.\cite{rabinovich2005capillary}    & $d=\frac{S}{2}(-1+\sqrt{1+2V_l/(\pi R_p S^2)} )$\\ 
    Model E
    & $F^* = \exp(A S^* + B) + C$\\
 Mikami et al.\cite{mikami1998numerical} &$A = - 1.1 \Bar{V}^{-0.53}$\\
&$B = (-0.34 \ln \Bar{V} - 0.96) \theta^2 - 0.019 \ln \Bar{V} + 0.48$\\
&$C = 0.0042 \ln \Bar{V} + 0.078$\\ 
&$\Bar{V}=\frac{4\pi}{3}V^*$\\
    \end{tabular}
\end{center}
\caption{Capillary force from different theoretical models (the
definitions of the coefficients $f_1, f_2, f_3$ and $f_4$ are
presented in the Supplementary Material). Note that the two expressions reported for Model B found in the literature are equivalent. In orginal models B (Megias \& Gauckler\cite{megias2010analysis}) and D, negative capillary forces indicate attraction whereas a positive force is repulsive. To keep consistency with our current study, we use positive capillary forces from models B and D for attraction.}
\label{table2}
\end{table*}

\subsection{Lattice Boltzmann Method}

The lattice Boltzmann method is a mesoscopic approach to approximate solutions of the Navier-Stokes equations by computing the moments of the Boltzmann transport equation solved on a lattice~\cite{benzi1992lattice}. Here, we model the droplet using 
the multicomponent lattice Boltzmann method of Shan and Chen~\cite{shan1993lattice,liu2016multiphase}. Each component follows a discretized Boltzmann equation
\begin{equation}\label{eq:3}
   f_i^c (\vec{x} + \vec{c}_i \Delta t, t+\Delta t ) = f_i^c (\vec{x}, t ) + \Omega_i^c (\vec{x}, t ),
\end{equation}
where $f_i^c (\vec{x}, t )$ represent the amount of particles of component $c$ at lattice position  $\vec{x}$  and at time $t$ that are moving along the $i$-th of the $N=19$ discretized directions, with velocity $\vec{c}_i (i = 1,\ldots,N)$ commensurate with the three-dimensional lattice. In reduced units, the timestep $\Delta t$ and the lattice constant $\Delta x$ are
set to 1. Here, $\Omega_i^c$ is the Bhatnagar-Gross-Krook (BGK) collision
operator representing the relaxation of the  particle distribution towards the local Maxwell-Boltzmann equilibrium~\cite{bhatnagar1954model}
\begin{equation}\label{eq:4}
    \Omega_i^c = -\frac{f_i^c (\vec{x}, t ) - f_i^{eq}[\rho^c(\vec{x},t), \vec{u}^c(\vec{x}, t) ] }{\tau^c},
\end{equation}
where the equilibrium distribution is approximated as
\begin{equation}\label{eq:5}
    f_i^{eq} = \zeta_i \rho^c \left[1 + \frac{\vec{c}_i\vec{u}}{c_s^2} + \frac{(\vec{c}_i\vec{u})^2}{2c_s^4} - \frac{u^2}{2c_s^2} \right].
\end{equation}
Here, $\tau^c$ sets the relaxation time. 
$\rho_c(\vec{x},t) = \rho_0 \sum_{i} f_i^c (\vec{x},t)$ is the fluid
density, and $\vec{u} =
\vec{u}^c(\vec{x},t)$ (defined by $\rho_c (\vec{x},t) \vec{u}^c (\vec{x},t) \equiv  \sum_i
f_i^c (\vec{x},t) \vec{c}_i$) is the macroscopic bulk velocity of the fluid. $\rho_0$ is a reference density and is chosen to be 1. $c_s = 1/\sqrt{3}$ is
the speed of sound and $\zeta_i$ are the coefficients from the velocity space discretization. $\nu_c
= c_s^2 (\tau_c - 1/2)$ is the kinematic viscosity of the fluid.

The fluid interaction between components $c$ and $c^\prime$ is
\begin{equation}\label{eq:6}
    \vec{F}^c (\vec{x}, t) \equiv - \Phi^c (\vec{x}, t)\sum_{c^{\prime}} G_{c c^{\prime}}\sum_{\vec{x}} \Phi^{c^{\prime}} (\vec{x}, t) (\vec{x}^{\prime}- \vec{x}),
\end{equation}
where $\vec{x}^\prime$ are the nearest neighboring lattice sites. $\Phi^c (\vec{x},t)$ is the effective mass and we use
\begin{equation}\label{eq:7}
    \Phi^c (\vec{x}, t) = \rho_0 \left[ 1 - \exp\left( - \frac{\rho^c(\vec{x}, t)}{\rho_0} \right) \right].
\end{equation}

$G_{c c^\prime}$ represents the coupling-constant of the interaction potential between components $c$ and $c^\prime$. In this work, we use $G_{c c^\prime} = 0.1$. The effect of the force is imposed by
adding a shift to the velocity $\vec{u}$ in the equilibrium distribution
\begin{equation}\label{eq:8}
    \Delta \vec{u}^c (\vec{x}, t) = \frac{\tau^c \vec{F}^c (\vec{x}, t)}{\rho^c(\vec{x}, t)}.
\end{equation}

The actual macroscopic bulk velocity is finally calculated as
\begin{equation}\label{eq:9}
    \rho^c(\vec{x}, t) \vec{u}^c (\vec{x}, t) = \sum_i f_i^c (\vec{x}, t) \vec{c}_i+ \frac{\vec{F}^c (\vec{x}, t)}{2}.
\end{equation}

The particles are discretized on the fluid lattice and interactions
between the fluid and the particles are introduced using a
modified bounce-back boundary condition,
resulting in a modified lattice Boltzmann equation~\cite{ladd1994numerical,aidun1998direct}. The particles follow Newton's equations of motion for the linear and angular momentum
\begin{equation}
    \vec{F}_p = m \frac{d \vec{u}_p}{dt}
\end{equation}
\begin{equation}
    \vec{D} = J \frac{d \vec{\omega}}{dt},
\end{equation}
where $\vec{F}$ is the total force acting on a particle, $m$ is the particle mass, and $\vec{u}_p$ its velocity. $\vec{D}$ is the torque, $J$ the particle's moment of inertia, and $\vec{\omega}$ its angular velocity.

The force in Eq.~\ref{eq:6} considers interactions between a lattice node outside of a particle and a lattice node inside a particle. To calculate these interactions, we fill the lattice nodes in the outer shell of the particle with a virtual fluid. We define a parameter
$\Delta \rho$, the particle colour, which allows us to control the
interaction between the particle surface and the two fluids as
\begin{equation}\label{eq:10}
 \rho_{virt}^r(\vec{x},t) = \rho^r(\vec{x},t)  + |\Delta \rho| 
\end{equation}
\begin{equation}\label{eq:11}
 \rho_{virt}^b(\vec{x},t) = \rho^b(\vec{x},t)  - |\Delta \rho|,
\end{equation}
where $\rho_{virt}^r(\vec{x},t)$ and $\rho_{virt}^b(\vec{x},t)$ are
the averages of the densities of neighboring fluid nodes for
components $r$ and $b$, respectively. A particle color $ \Delta
\rho= 0$ corresponds to a contact angle of $\theta = 90^\circ$, i.e.,
a neutrally wetting particle~\cite{JH11,DKCH14}.

%We perform simulations using a simulation box size of $256 \times 256 \times 448$ lattice units%(A box size of $512 \times 512 \times 512$ is tested, but almost the same capillary bridge is obtained). 

We place two particles ($R_p=80$ lattice units for equal spheres, $R_p=70$ and $90$ lattice units, respectively, for unequally sized spheres) along the $x$ axis, separated by a surface-to-surface distance $S$. A droplet is initialized in the center of the system.
The droplet size is chosen large enough to make sure that the capillary
bridge can form between two particles (a minimum droplet radius of 23 lattice sites is used). As constant input parameters, we prescribe the
liquid bridge volume V$_l$ = V$_{bridge}$, the contact angle $\theta$,
the particle-particle separation distance $S$ and the surface
tension of the fluid, $\sigma$. By changing the droplet radius and length, we can
obtain a wide range of liquid-to-solid volume ratios. We fix the
position and orientation of the particles and let the system
equilibrate. Then, we measure the forces acting on the
particles for different values of the separation, contact angle, and volume ratio. To study the effect of discretization artefacts, we carried out a convergence study for the capillary force calculation, as a function of the grid size. In these tests, the relative liquid bridge volume is fixed at 0.133, and the effective resolution is increased by employing larger particle radii and correspondingly a higher number of lattice sites. Table~\ref{table3} summarizes the calculated capillary force. By increasing the resolution by a factor of three, the change in force is only about 2.5\%. This value can be therefore used as an estimate of the (quite small) discretization error.

\begin{table*}
\setlength\arrayrulewidth{1.0pt}
\begin{center}
    \begin{tabular}{llll}
    \hline
    Box size & & & $F/(\sigma R_p)$\\ 
    \hline
    128$\times$128$\times$248 & & & 1.92\\
    192$\times$192$\times$336 & & & 1.94\\
    256$\times$256$\times$448 & & & 1.94\\
    384$\times$384$\times$608 & & & 1.97\\
  \hline
    \end{tabular}
\end{center}
\caption{Dimensionless capillary force ($F/(\sigma R_p)$) calculated from LBM-DEM at $R_p/N_x = 0.3125$, $S/R_p=0.625$ and $V_l/V_p = 0.133$.}
\label{table3}
\end{table*}

If not specified otherwise, we present all results in dimensionless units.
The dimensionless force, bridge volume and separation distance are
defined as $F^*=\frac{F}{\sigma R_p}$, $V^*=\frac{V_l}{V_p}$, and
$S^*=\frac{S}{R_p}$.

\section{Results and Discussion} 
\begin{figure*}[tbp]
\centering
\subcaptionbox{$V^* = 0.07$\label{fig:2a}}{\includegraphics[width=0.6\columnwidth]{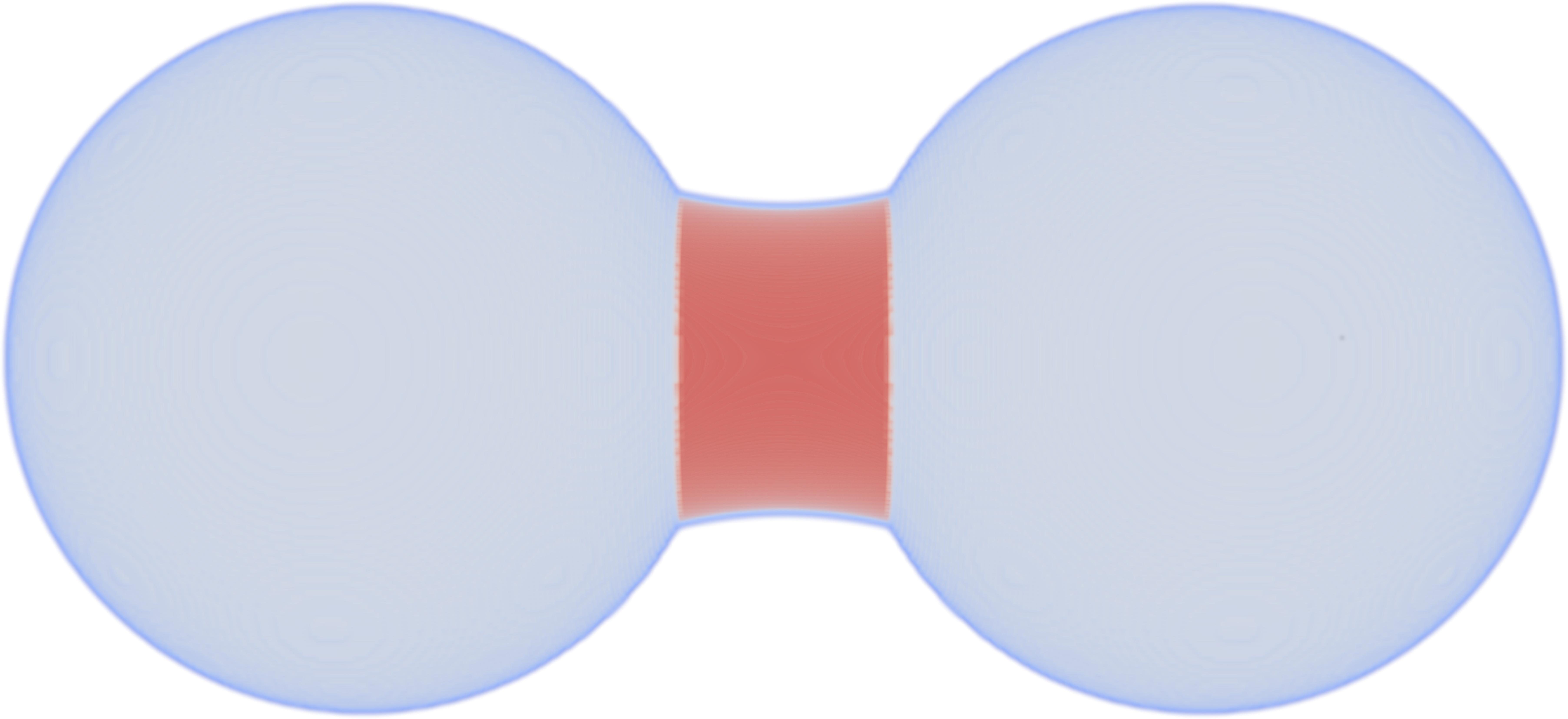}}\hspace{.3em}%
\subcaptionbox{$V^* = 0.133$\label{fig:2b}}{\includegraphics[width=0.6\columnwidth]{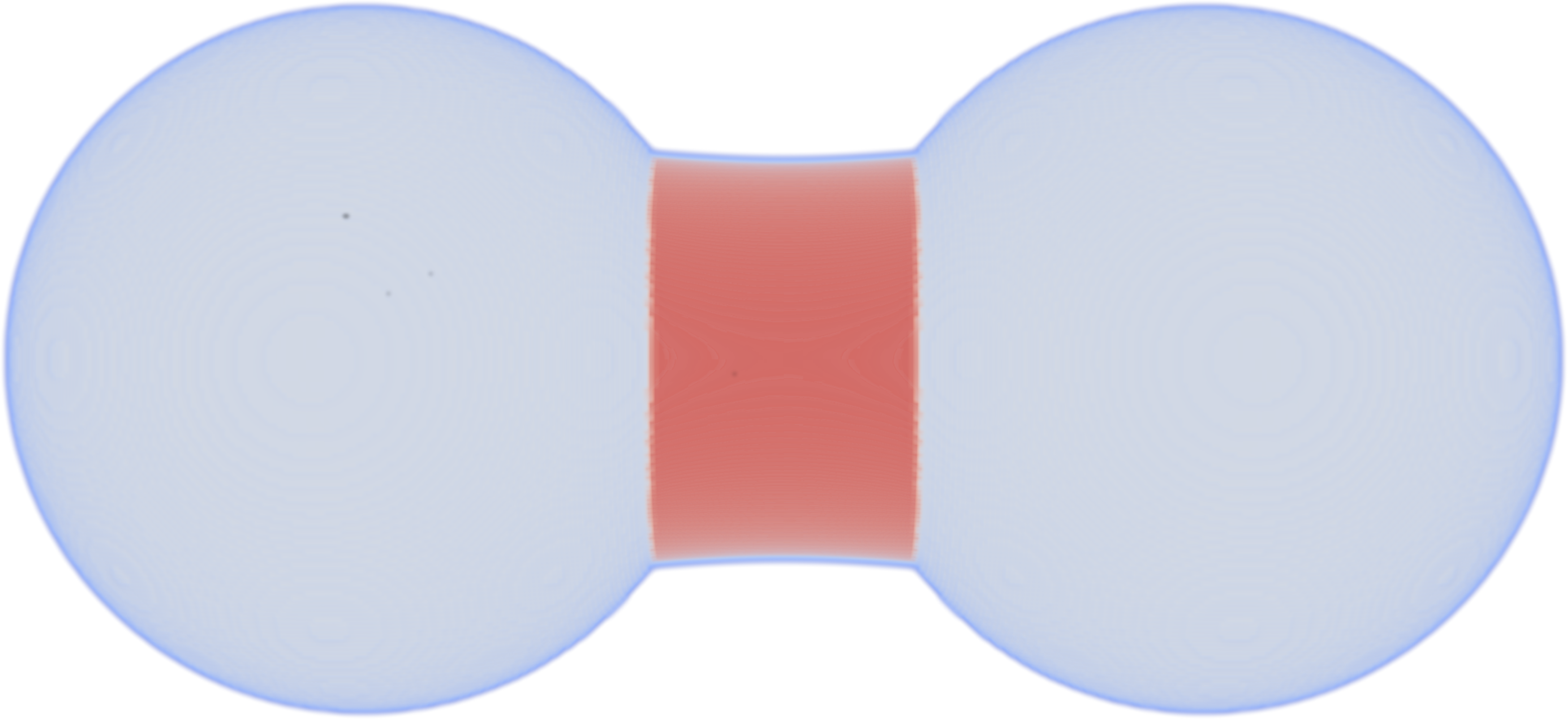}}
\subcaptionbox{$V^* = 0.25$\label{fig:2c}}{\includegraphics[width=0.6\columnwidth]{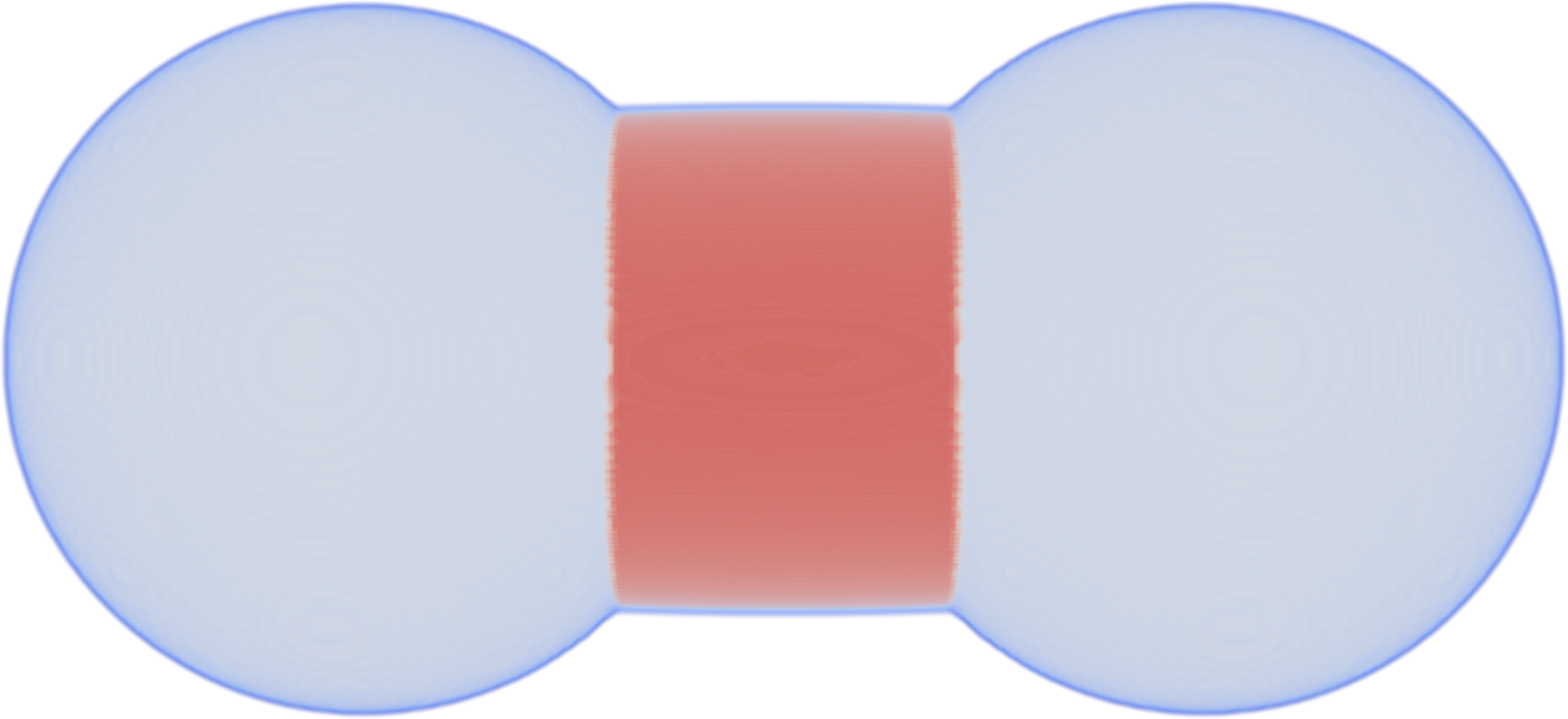}}\\
\subcaptionbox{contact angle=34.6$^\circ$\label{fig:5d}}{\includegraphics[width=0.6\columnwidth]{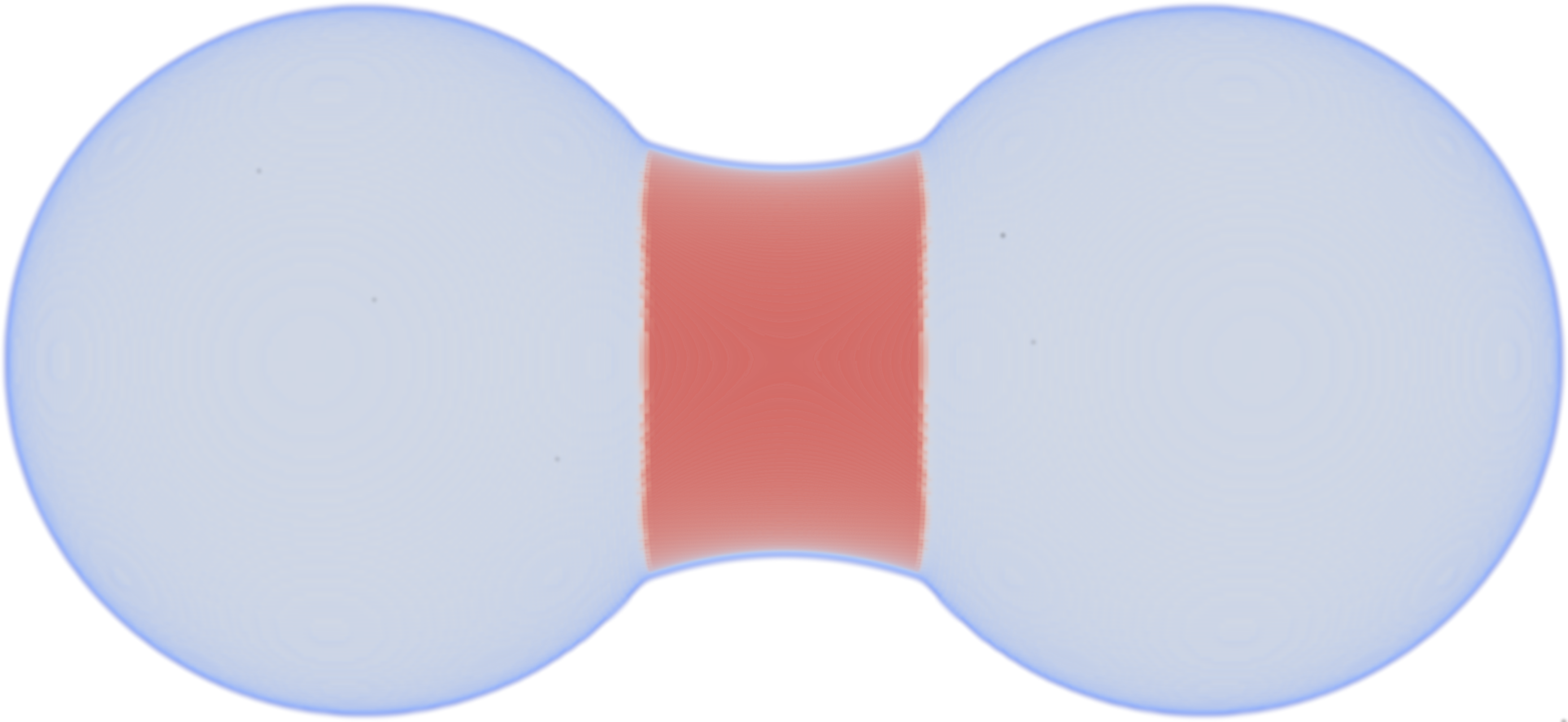}}
\subcaptionbox{contact angle=60$^\circ$\label{fig:5e}}{\includegraphics[width=0.6\columnwidth]{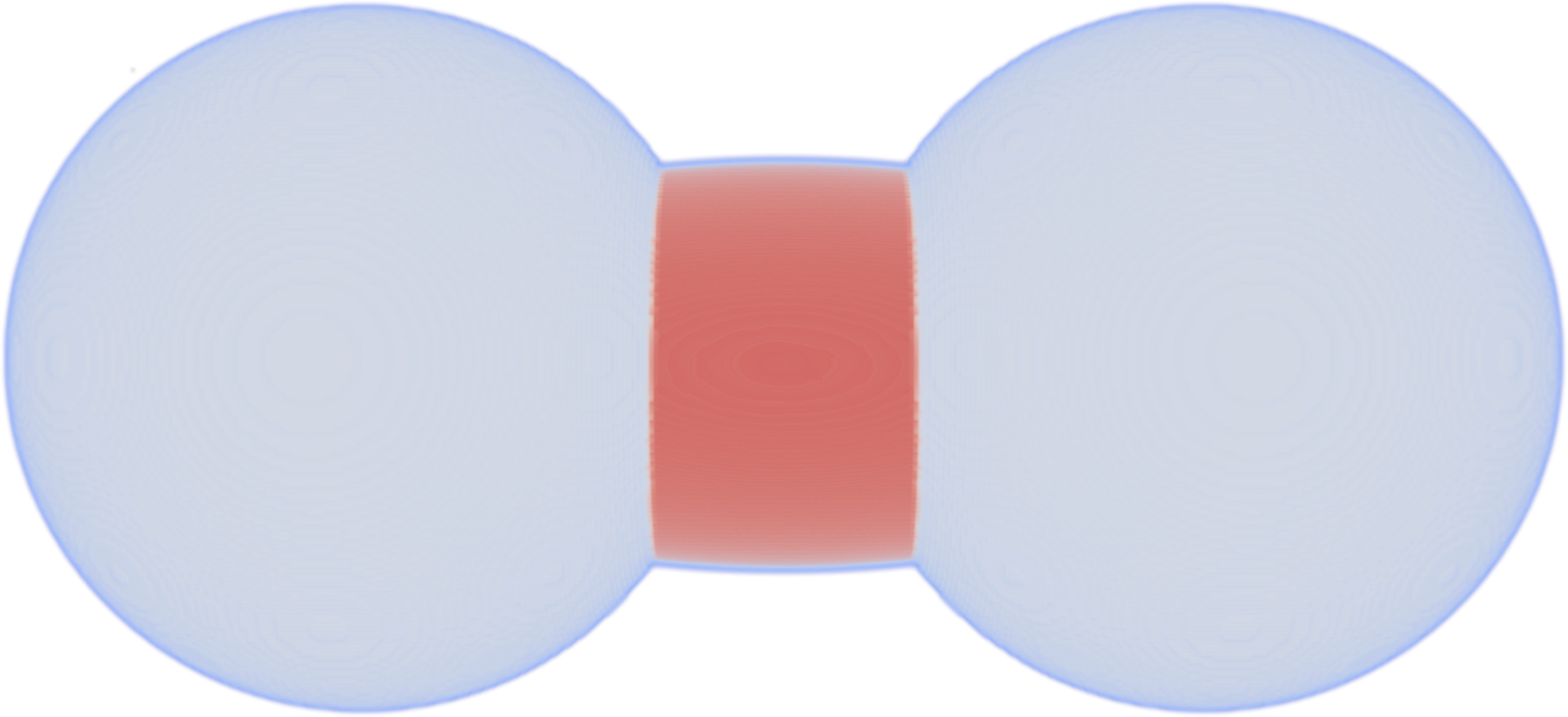}}
\caption{Liquid bridge profile with different volume ratios under fixed separation distance ($S/R_p=0.375$) (top row: contact angle =$48^\circ$) and contact angles (bottom row: volume ratio $V^*=0.133$).
}
\label{fig:snap1}
\end{figure*}

We vary the ratio of bridge volume to particle volume $V^*$($V_l/V_p$) 
from 0.07 to 0.7. The dimensionless separation distance $S^*$($S/R_p$) between spheres
is changed from 0.25 to 1.45 to obtain the corresponding liquid bridge force. The influence of the contact angle is tested by
performing simulations for $\theta=34.6^\circ, 48^\circ, 60^\circ$ and $74^\circ$,
respectively.

We report the bridge profiles under different liquid volumes and contact
angles in Fig.~\ref{fig:snap1} using the same liquid volume $V^*=0.133$. As expected, the bridge shape is symmetric. However, one can appreciate that the meniscus changes from concave to convex when the liquid volume is increased. This change is induced by the geometric constraints, even though the wetting parameter (defining the contact angle of a droplet on a plane surface) is kept constant.

\begin{figure*}
\centering
        \includegraphics[width=1.0\textwidth]{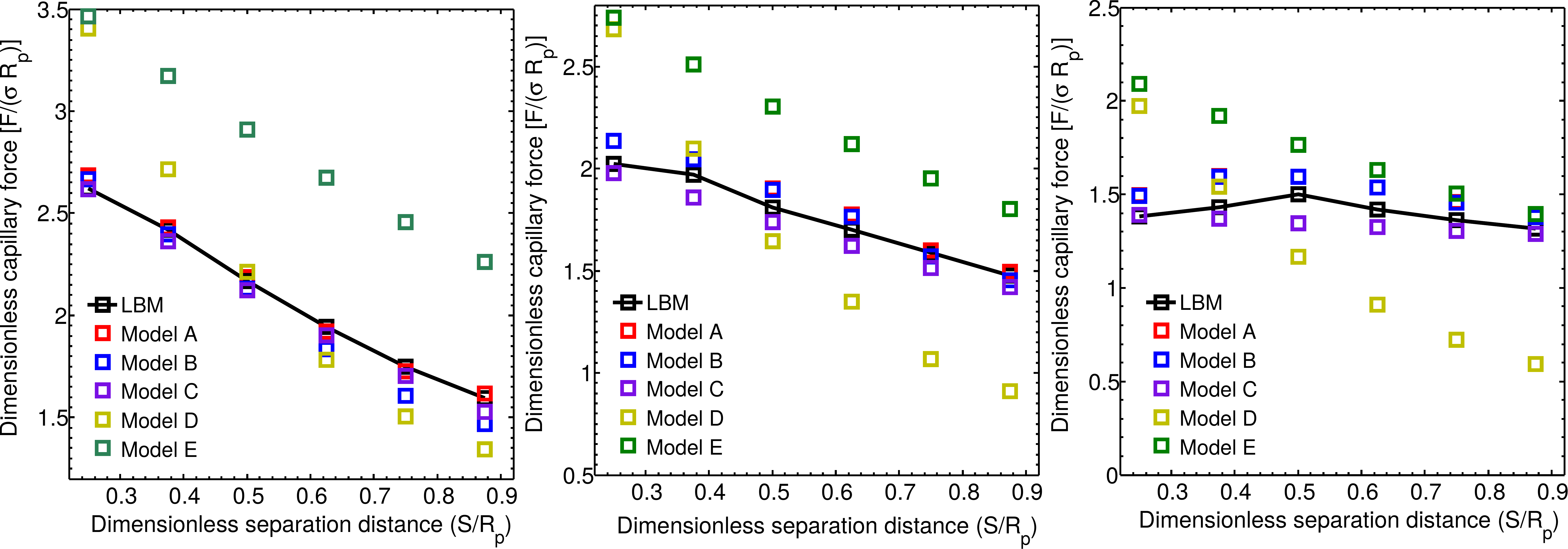}
\caption{Capillary bridge force between two equally-sized spheres with bridge volume $V^*=0.133$, for different contact angles (left panel: $34.6^\circ$, middle panel: $48^\circ$, right panel: $60^\circ$).}
  \label{fig:force-small-volume}
\end{figure*}
In Fig.~\ref{fig:force-small-volume}  we compare the force/distance relation obtained with the lattice Boltzmann method to the predictions of the theoretical models at various contact angles. A similar behaviors are observed at different bridge volumes, and reported in the Supplementary Material).

Models A, B and C show a very good qualitative and quantitative agreement with the simulation results, with relatively small discrepancies ($<10\%$), nicely reproducing the different trends characterizing the different contact angles. Models A and B predict slightly higher capillary forces than calculated with the lattice Boltzmann simulation, while model C predicts slightly lower values. Models D and E are both qualitatively and quantitatively not satisfactory, showing a slope of the force/distance curve that is only mildly dependent on the contact angle. While Model D partly underestimates and partly overestimates the capillary bridge force, Model E constantly overestimates it by a factor 1.5 to 3. 

\begin{figure*}[tbp]
  \begin{subfigure}[b]{0.45\textwidth}
    \includegraphics[scale=0.45,width=2.0in]{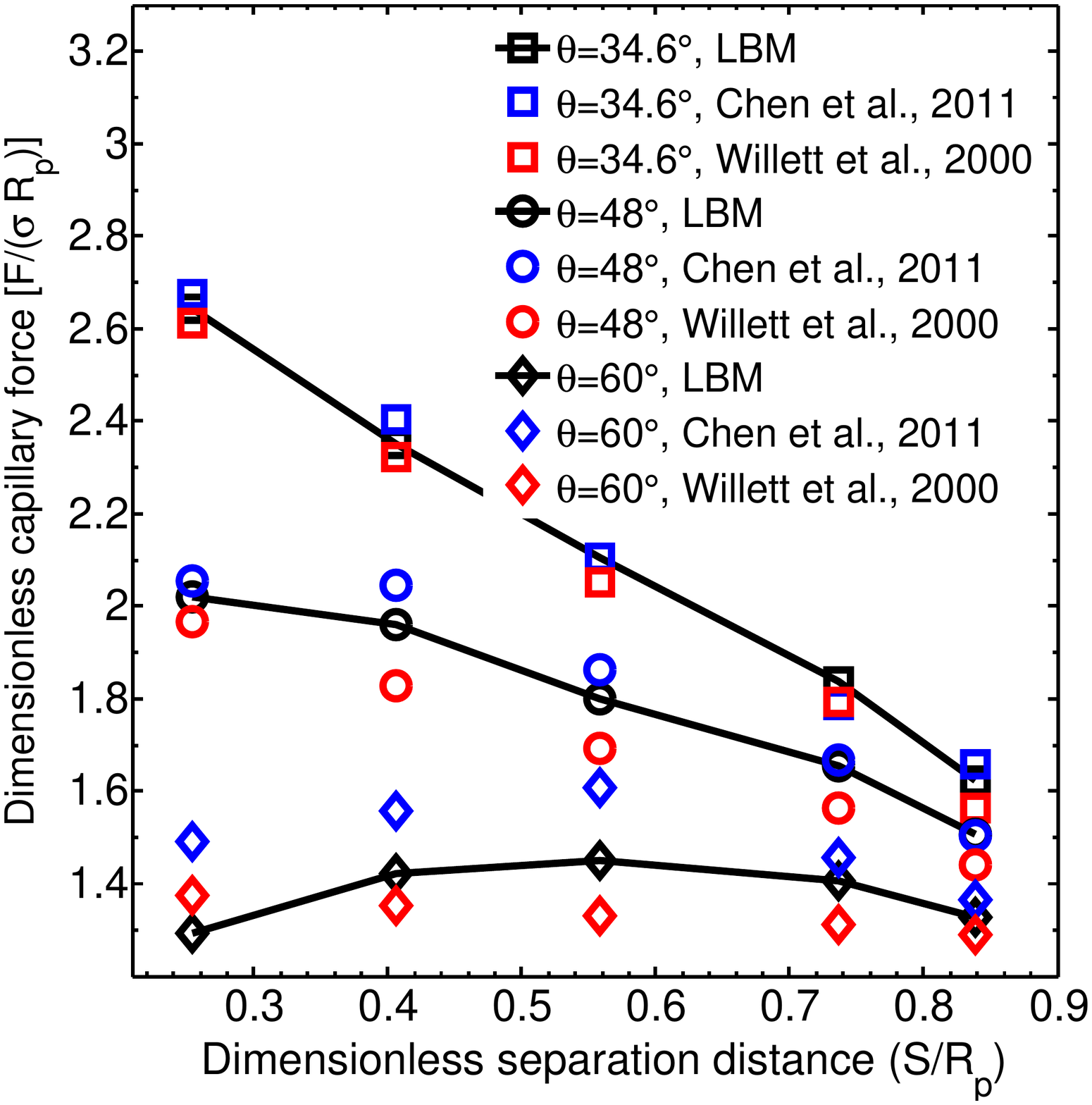}
       \caption{Bridge volume $V^*=0.07$} 
  \end{subfigure} 
  \hspace{3mm}
   \begin{subfigure}[b]{0.45\textwidth}
    \includegraphics[scale=0.45,width=2.0in]{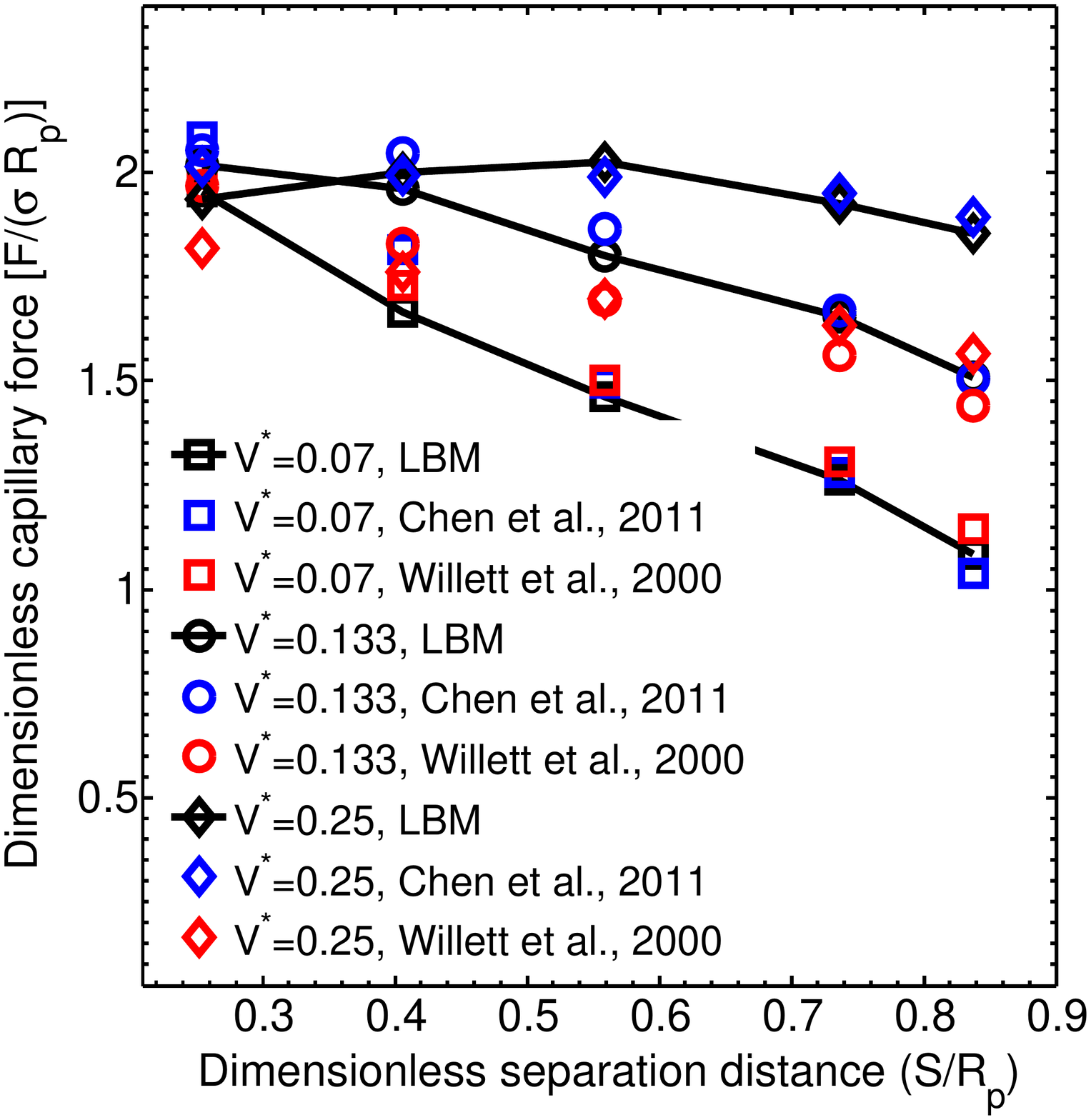}
        \caption{Bridge volume $V^*=0.25$} 
    \end{subfigure}
  \caption{Dimensionless capillary bridge force ($F/(\sigma R_p)$) between two equal spheres under different bridge volume. (Contact angle $\theta=48^\circ$)}
\label{fig:force-large-volume}
\end{figure*}

The overall trend indicates the capillary force decreasing with increasing particle
separation, due to the decreasing surface tension force.
When the contact angle or the bridge volume increase, the slope of the force/distance curve decreases. And at  $\theta=60^\circ$ or $V^*=0.25$ (see Fig \ref{fig:force-large-volume}b), the capillary force/distance curve becomes almost flat, except for the appearance of a maximum at a reduced separation distance $S/R_p\simeq0.5$. In this case, the bridge profile is convex at small separation $S$ while it becomes concave at large $S$. The peak displayed by the capillary force, previously already reported~\cite{tselishchev2003influence,megias2009capillary}, originates from the non-monotonic Laplace component~\cite{megias2009capillary}.

%%%%%%%%%%%% HERE
Next, we present results on capillary bridges between two static unequally sized spheres with identical contact angles. 
There is limited work in the literature regarding spheres of different size. Chen and coworkers (and later Sun and Sakai~\cite{sun2018liquid}) proposed a mechanical model to analyze the force and the volume of the liquid bridge, by considering a circular arc-shaped liquid bridge profile between two unequally sized
spheres~\cite{chen2011liquid}. In the following, we denote these models as Model F, and we provide a comparison of its predictions with the results of our lattice Boltzmann simulations. 
In addition, we show the predictions of Model C, computed using the harmonic mean of the radii $R_m=\frac{2R_1 R_2}{R_1+R_2}$ to calculate
capillary forces and bridge volumes~\cite{willett2000capillary}. In the simulations and model F, we use $R_p =$ 70 and 90 lattice units, respectively.

\begin{figure*}[tbp]
\begin{center}
\includegraphics[width=0.40\textwidth]{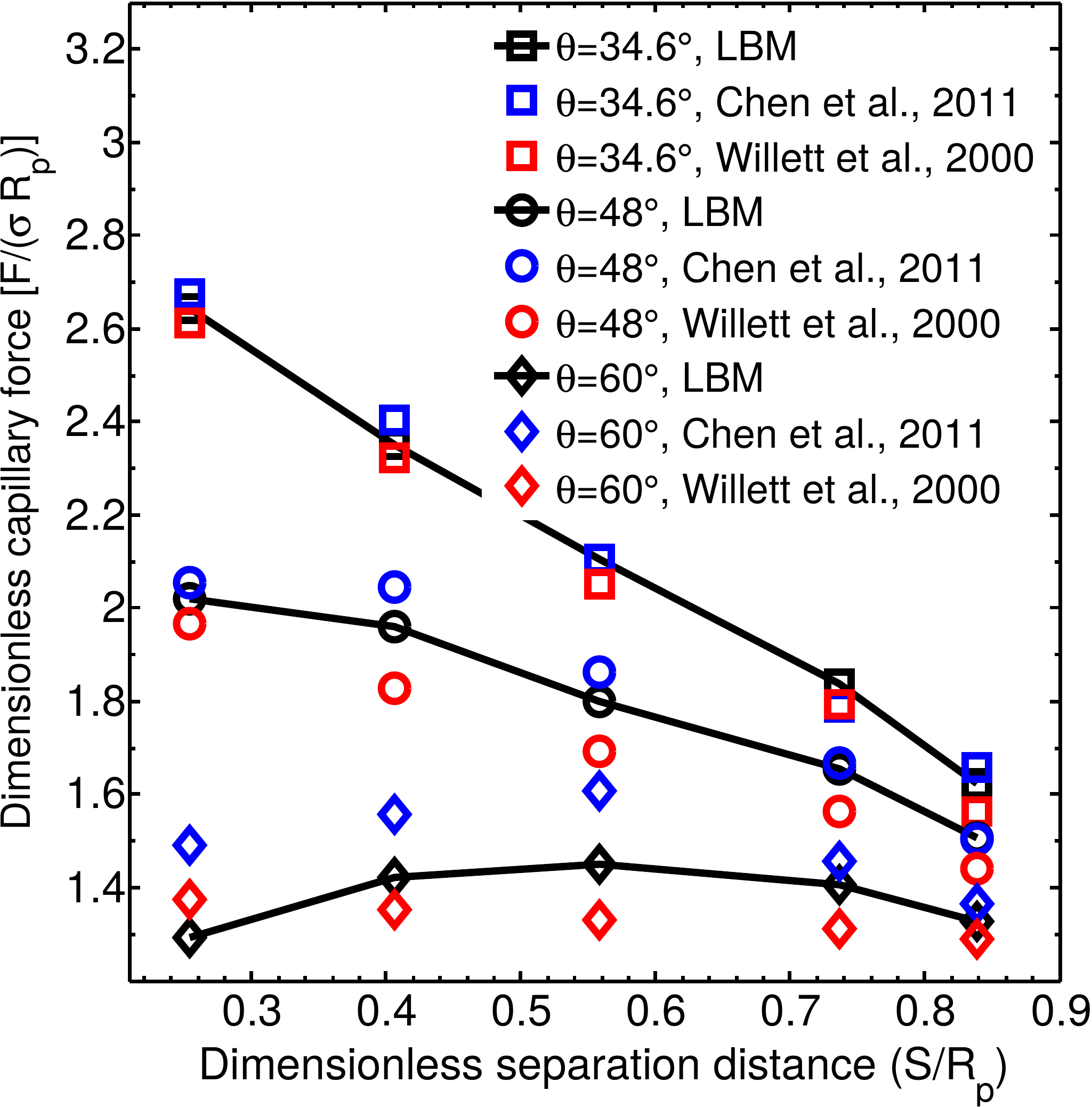}\hspace{1em}%
\includegraphics[width=0.40\textwidth]{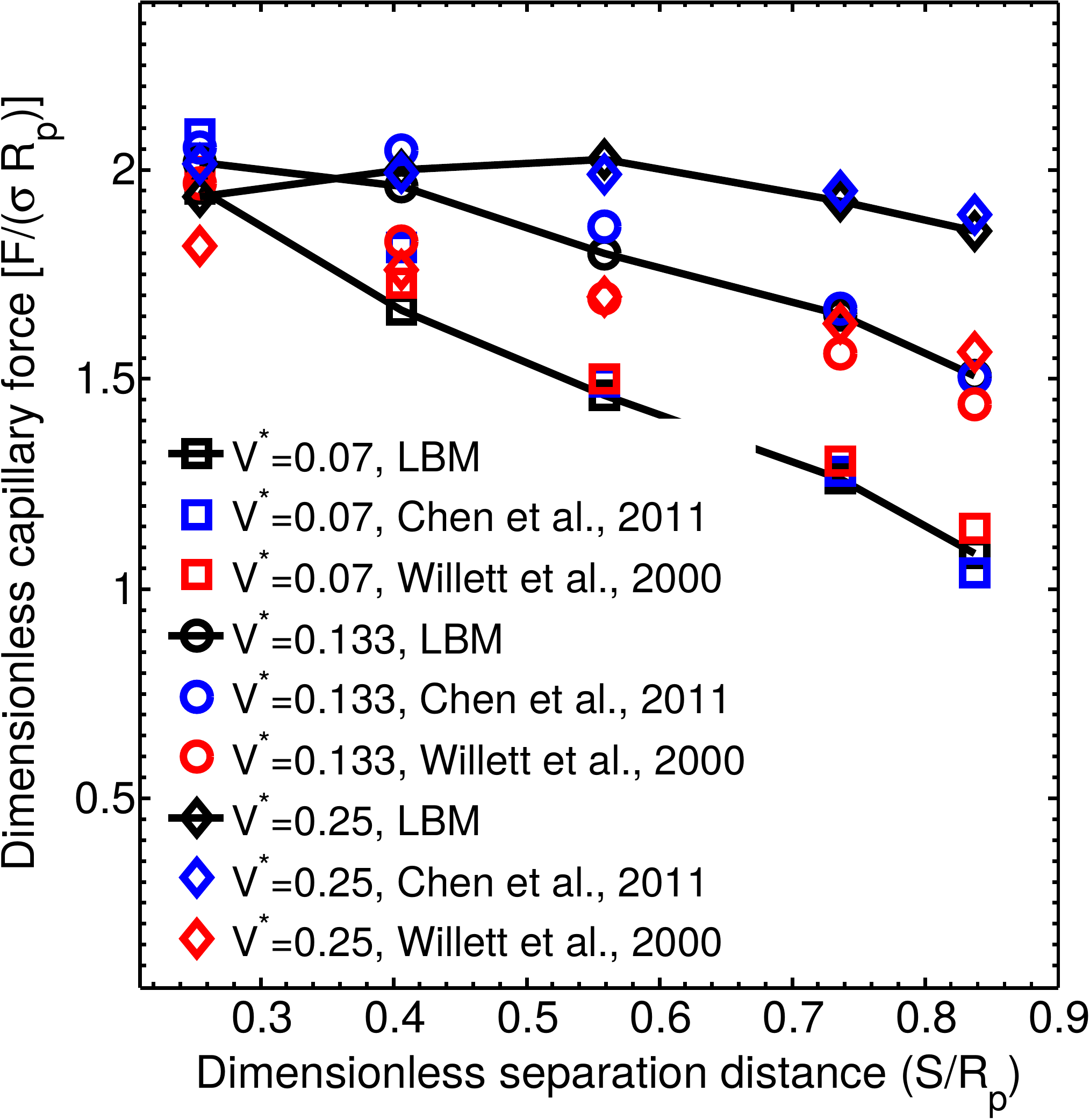}
\end{center}
       \caption{Capillary bridge force between two unequally-sized spheres with radii equal to 70 and 90 lattice units, respectively. Left panel: different contact angles and bridge volume $V^*=0.133$. Right panel: different bridge volumes and the same contact angle $\theta = 48^\circ$.
}
\label{fig:force-unequal-spheres}
\end{figure*}

In Fig~\ref{fig:force-unequal-spheres} we report the force/distance curves for unequal spheres. The behavior is qualitatively similar to the equal sphere case. At moderate contact angles and liquid volumes, the capillary force decreases with increasing particle separation due to the decreasing surface tension force, and a maximum in the force appears for the largest contact angle and liquid volume considered, much like in the symmetric system.  

Model F captures qualitatively very well the trend of the force/distance curves, also at high contact angles and liquid bridge volumes, and is also quantitatively accurate under most conditions.
Model C is worse than Model F in reproducing the qualitative trend at high contact angles and high liquid bridge volumes. However, from the quantitative point of view, Model C and Model F provide, on average, comparable predictions.

In the Supplementary Material, we also report the predictions of Models C and F, where we use, again, the harmonic mean of the radii.

\begin{figure*}[tbp]
\begin{center}
\begin{subfigure}[b]{0.45\textwidth}
    \includegraphics[scale=0.45,width=1.6in]{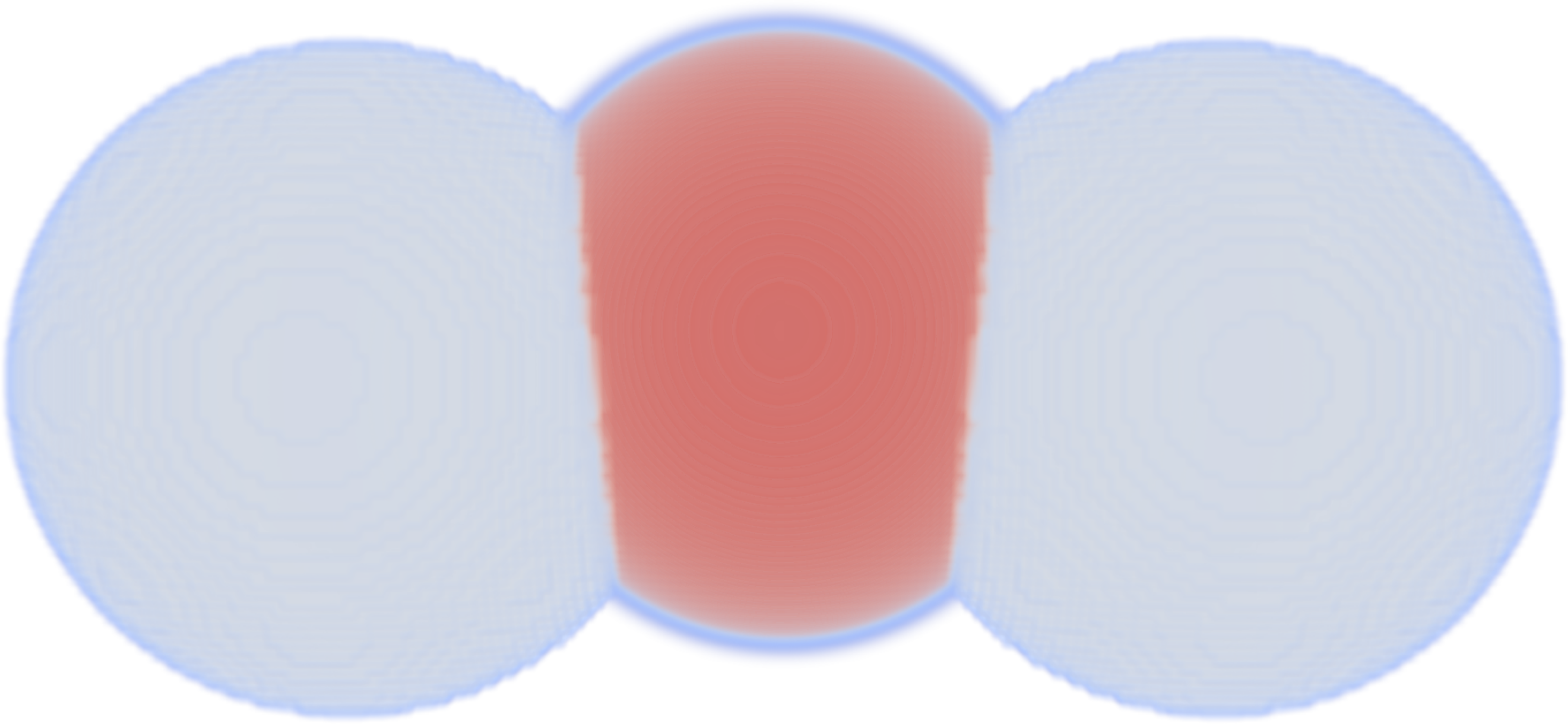}
       \caption{(a) $R^*=0.572$, $\theta = 90^\circ$} 
  \end{subfigure} 
  \hspace{1em}%
   \begin{subfigure}[b]{0.45\textwidth}
    \includegraphics[scale=0.45,width=1.6in]{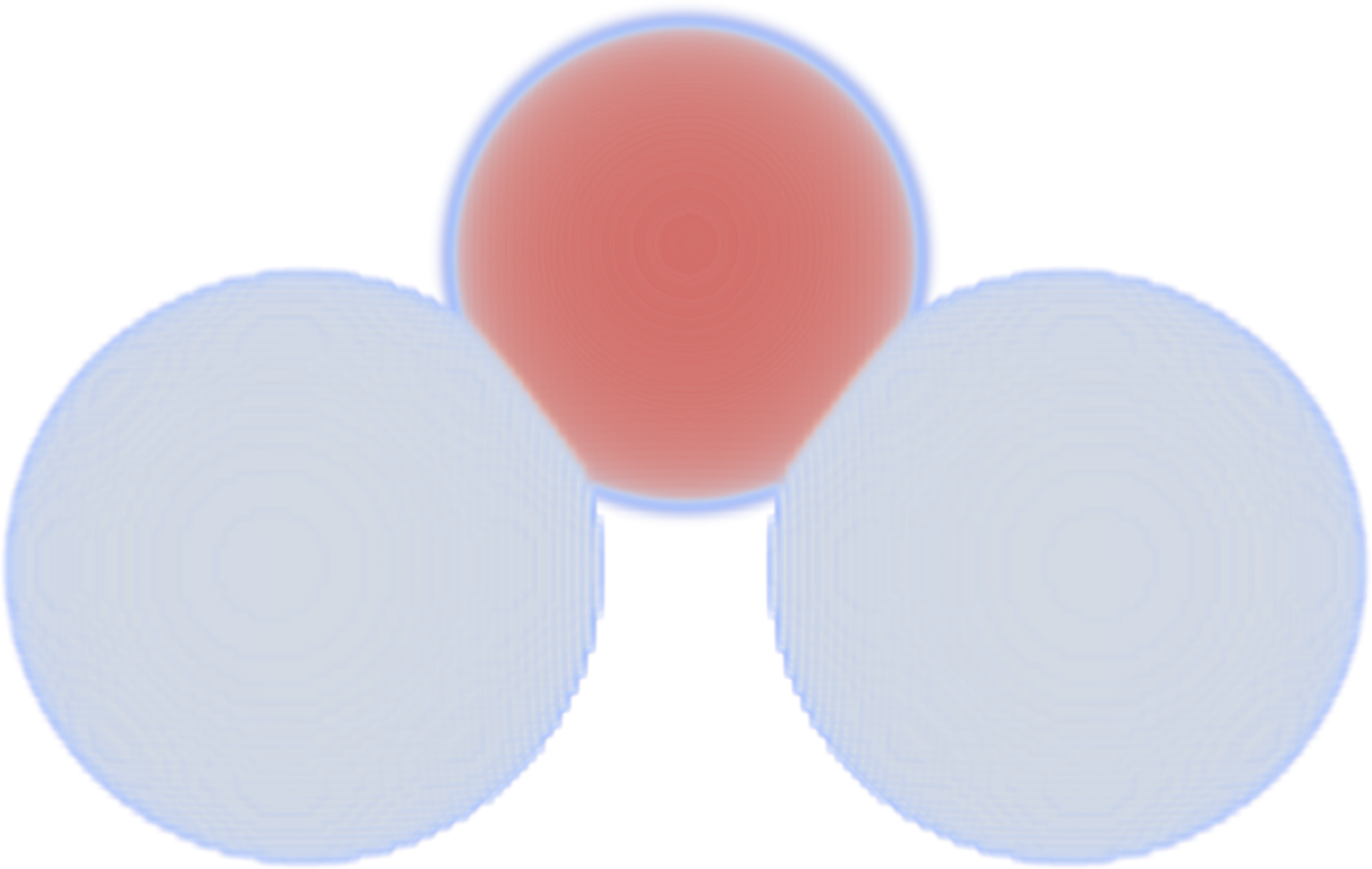}
        \caption{(b) $R^*=0.572$, $\theta = 132^\circ$ } 
    \end{subfigure}\\
\begin{subfigure}[b]{0.45\textwidth}
    \includegraphics[scale=0.45,width=1.8in]{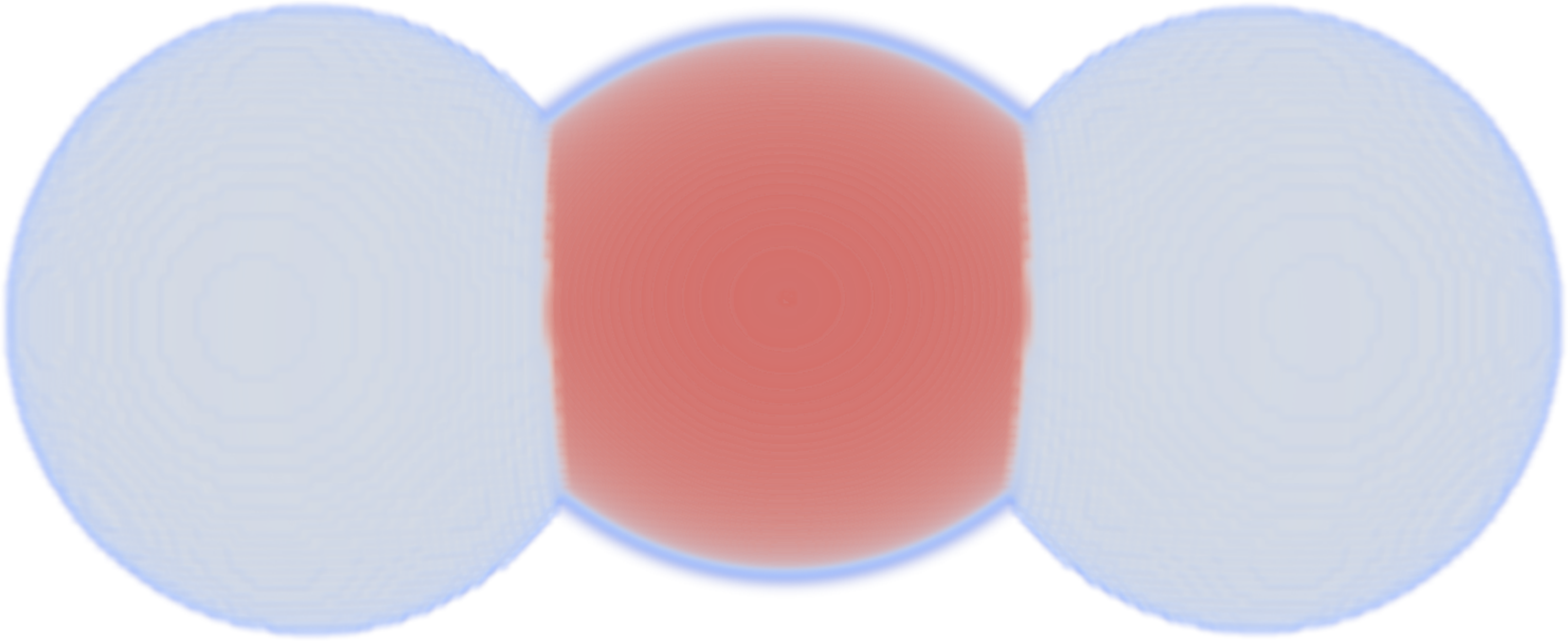}
       \caption{$R^*=1.144$, $\theta = 90^\circ$} 
  \end{subfigure} 
  \hspace{1em}%
   \begin{subfigure}[b]{0.45\textwidth}
    \includegraphics[scale=0.45,width=1.8in]{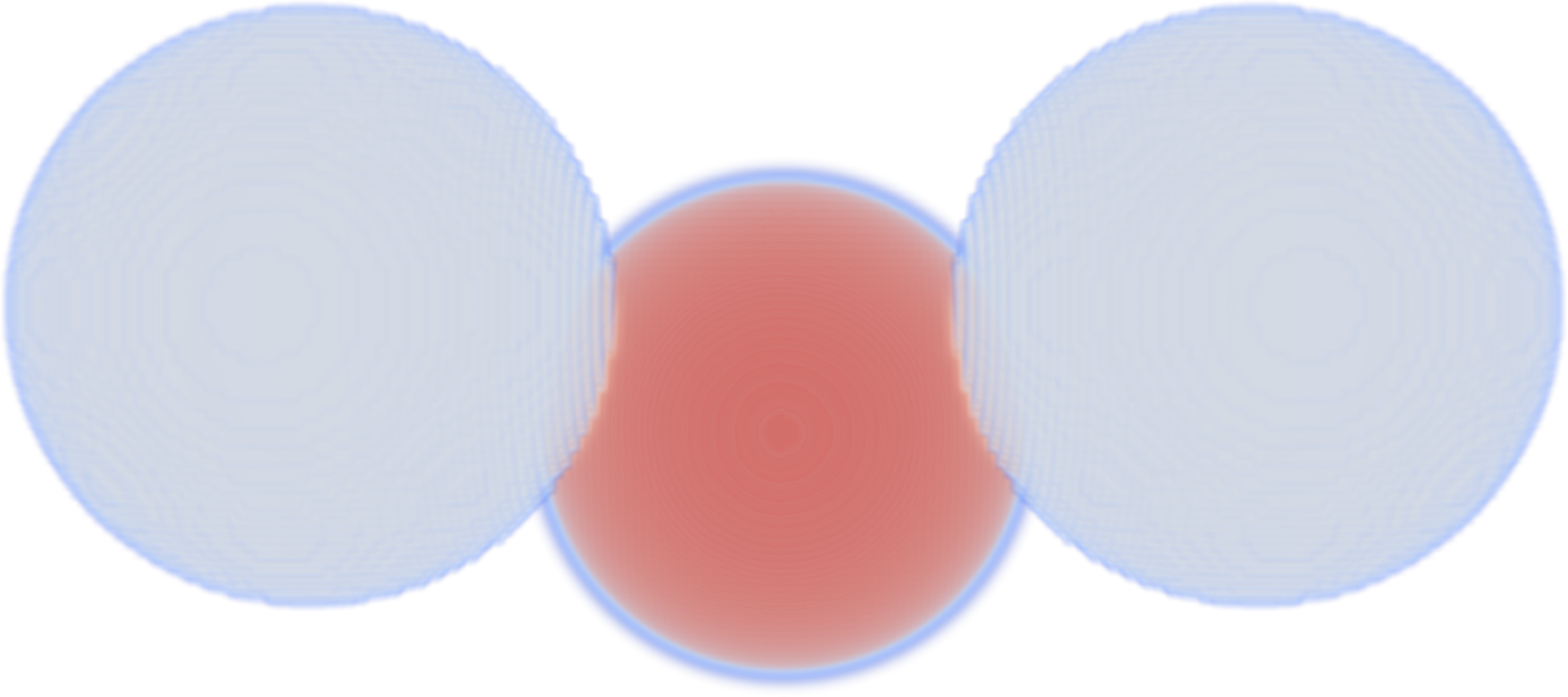}
        \caption{(d) $R^*=1.144$, $\theta = 132^\circ$} 
    \end{subfigure}
\end{center}
  \caption{Asymmetric liquid bridge between equal spheres from lattice Boltzmann simulation at different separation distances ($V^*=0.5$ for $S^*=0.572$, $V^*=0.65$ for $S^*=1.144$)}
\label{fig:asymm}
\end{figure*}

The results presented above show that several models have problems in reproducing the force/separation curves when the liquid bridge volume starts becoming comparable with that of the solid particles. The capillary bridge volume has an influence on the bridge shape (and consequently on the force) has also been demonstrated by experiments~\cite{farmer2015asymmetric}.
In fact, at high volumes or contact angles, as reported by Niven, liquid bridges tend even to adopt non-axisymmetric geometries~\cite{niven2006force}.
The lattice Boltzmann simulations can also reproduce this condition. Fig.~\ref{fig:asymm} illustrates the appearance of an asymmetric liquid bridge (e.g. tear-drop shape) between
spheres at different contact angles and bridge volumes. A droplet sits initially between two fixed particles. Then it migrates from its axisymmetric configuration to one side due to the pressure gradient produced from the top and bottom curvatures. In the case of the highest contact angle ($132^\circ$) a mechanically stable state is reached, and the capillary force is zero, whereas for neutral wetting conditions ($\theta=90^\circ$), we measure non-zero attractive forces. Clearly, these configurations cannot be described by the models discussed here. Interestingly, however, models A, B, C and D would predict a transition from attractive to repulsive capillary forces.  

\subsection*{Many particles system under shear flow}

\begin{figure*}[tbp]
\begin{center}
 %\captionsetup{justification=raggedright}
  \begin{subfigure}[b]{0.45\textwidth}
    \includegraphics[width=0.8\columnwidth]{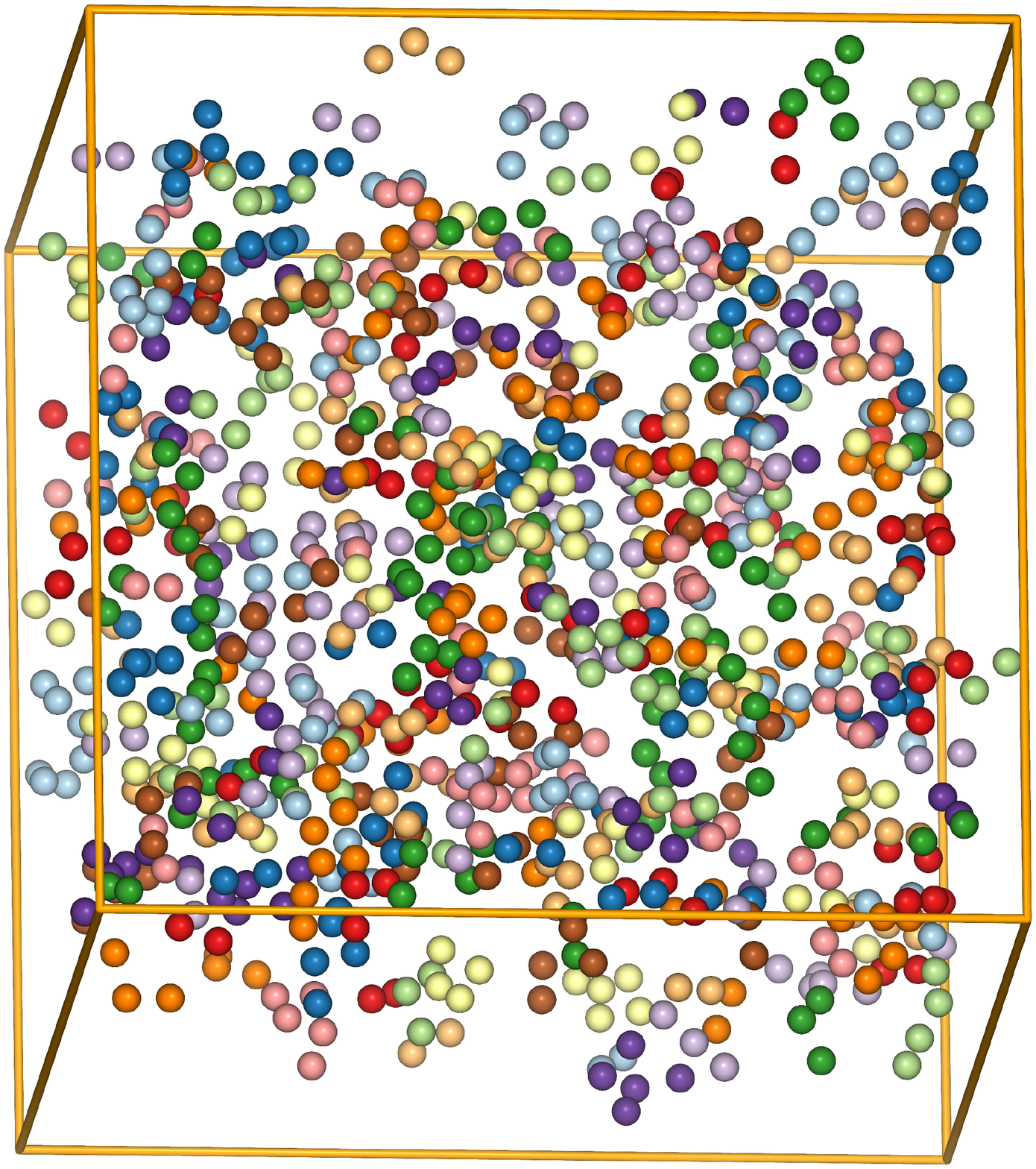}
       \caption{No capillary force} 
  \end{subfigure}
  \hspace{3mm}
   \begin{subfigure}[b]{0.45\textwidth}
    \includegraphics[width=0.8\columnwidth]{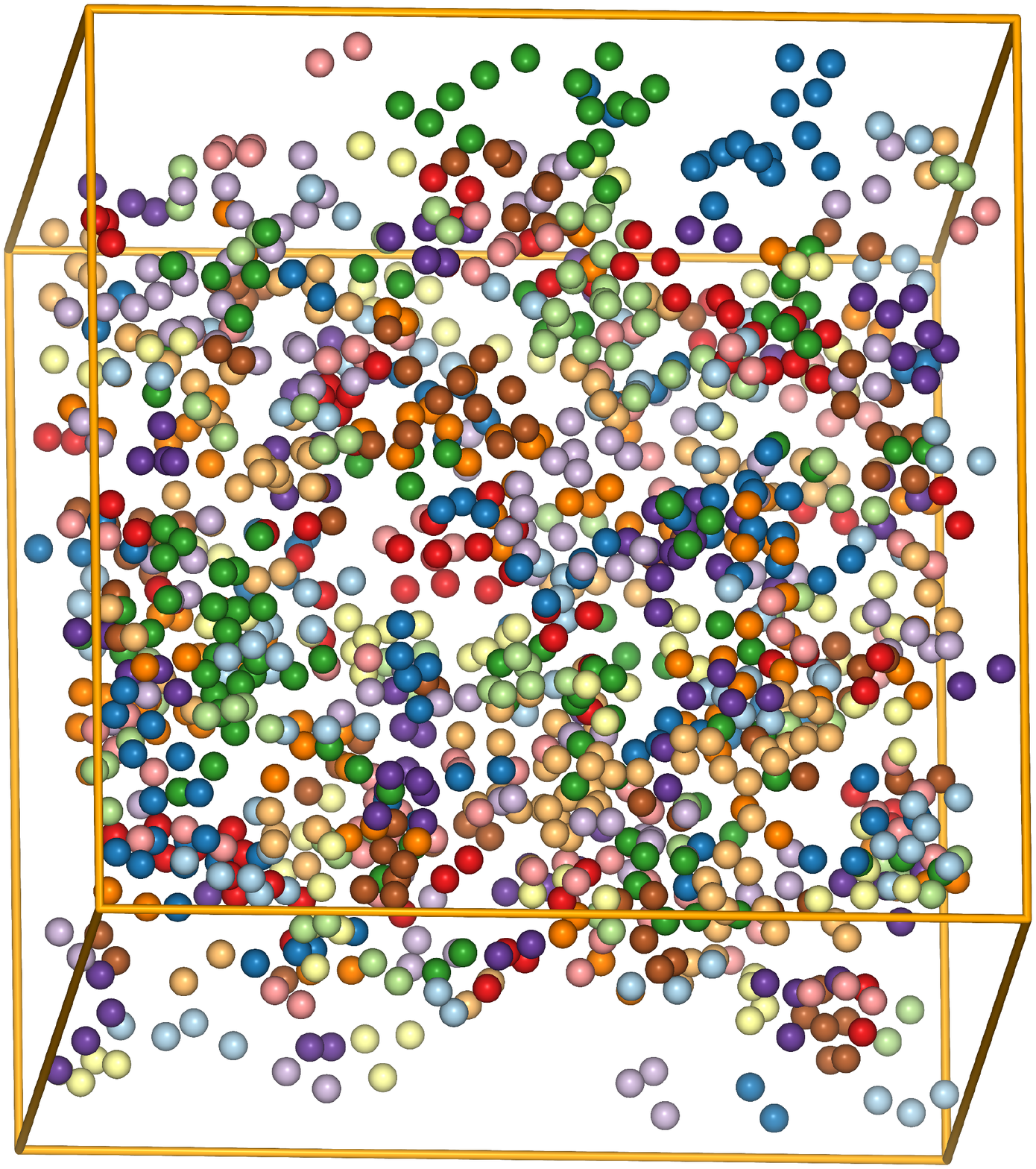}
      \caption{$\sigma=1\times10^{-5}$}
    \end{subfigure} \\
  \begin{subfigure}[b]{0.45\textwidth}
    \includegraphics[width=0.8\columnwidth]{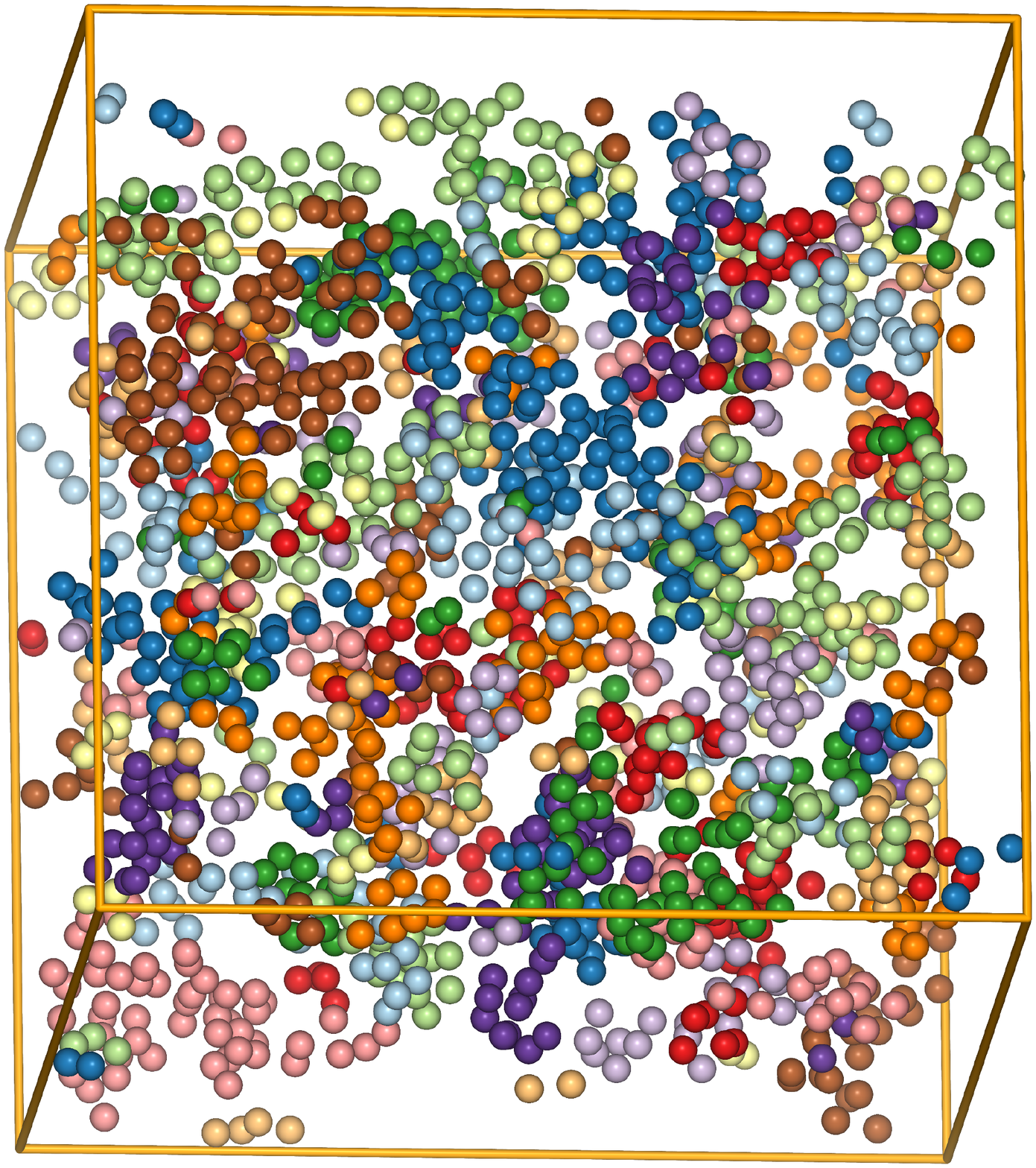}
    \caption{$\sigma=5\times10^{-5}$} 
  \end{subfigure}
  \hspace{3mm}
   \begin{subfigure}[b]{0.45\textwidth}
    \includegraphics[width=0.8\columnwidth]{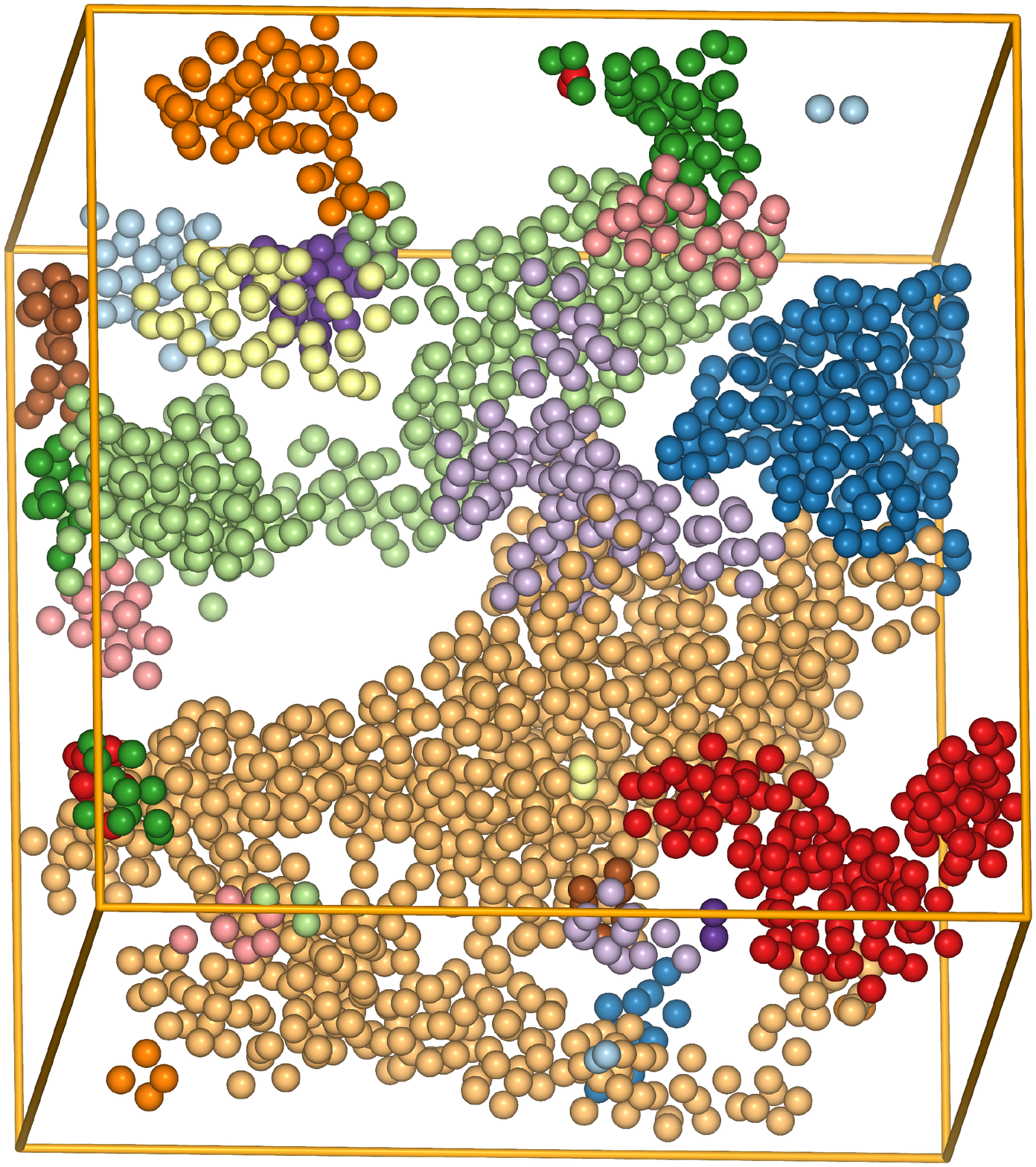}
      \caption{$\sigma=5\times10^{-4}$} 
    \end{subfigure} 
\end{center}
    \caption{Simulation snapshots of particle clustering at $t=4 \times 10^6 \Delta t$ under different capillary forces in a shear flow. Particles  belonging to the same cluster have the same color. (a: no capillary force; b,c,d: capillary force with different surface tension parameter $\sigma$).}
\label{fig8}
\end{figure*}

\begin{figure*}[tbp]
%\begin{center}
 %\captionsetup{justification=raggedright}
  \begin{subfigure}[b]{0.45\textwidth}
    \includegraphics[scale=0.45,width=2.8in]{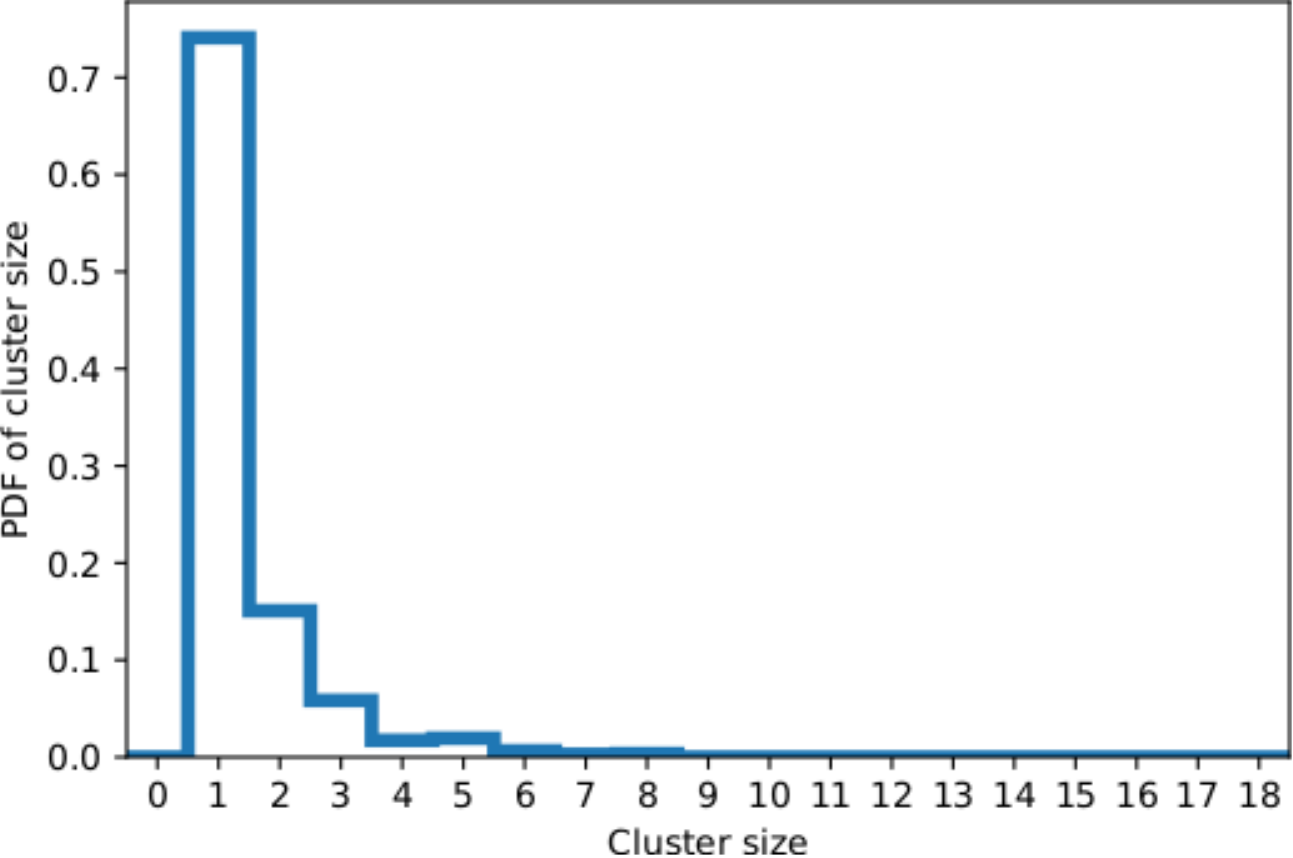}
       \caption{No capillary force} 
  \end{subfigure}
  \hspace{3mm}
   \begin{subfigure}[b]{0.45\textwidth}
    \includegraphics[scale=0.45,width=2.8in]{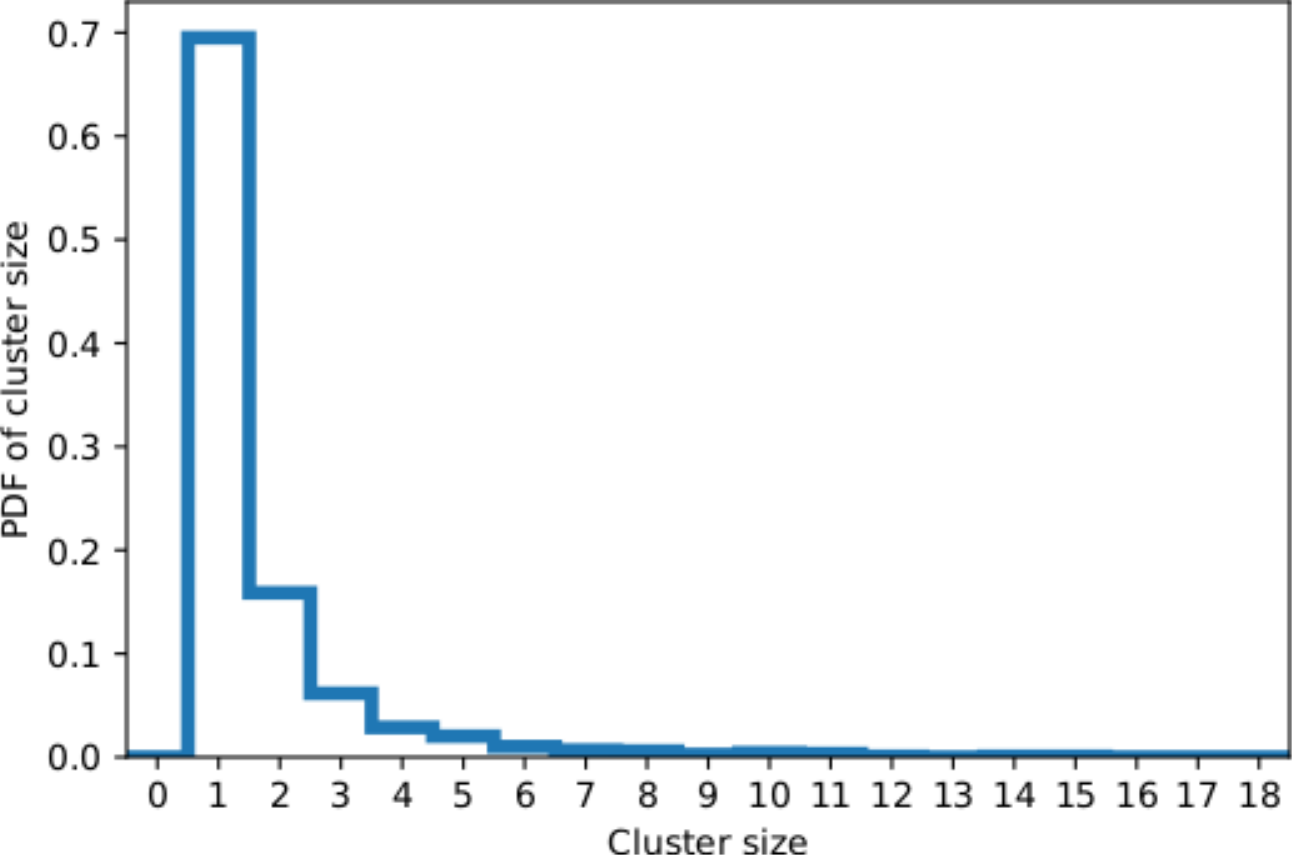}
     \caption{$\sigma=10^{-5}$}
    \end{subfigure} \\
\begin{subfigure}[b]{0.45\textwidth}
    \includegraphics[scale=0.45,width=2.8in]{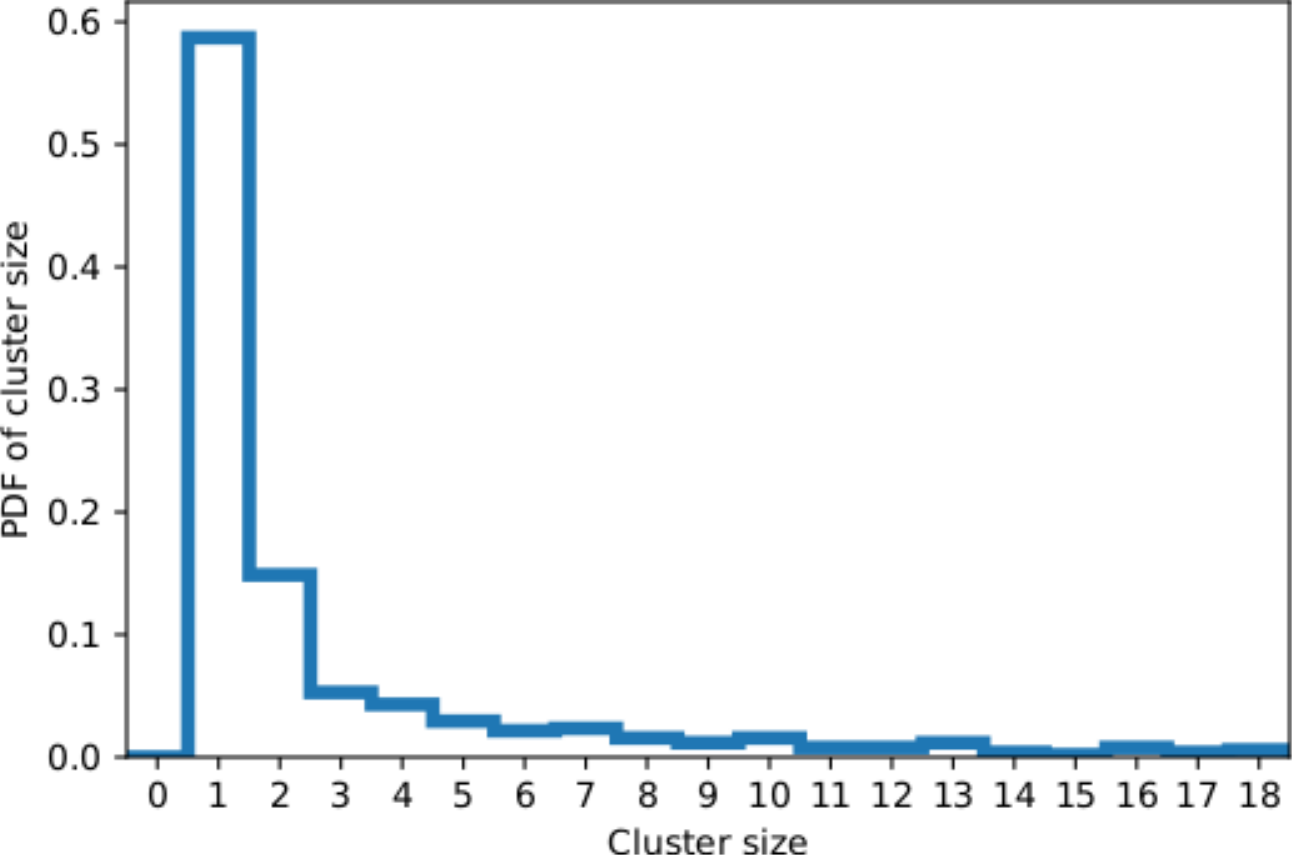}
      \caption{$\sigma=5\times10^{-5}$} 
  \end{subfigure}
  \hspace{3mm}
   \begin{subfigure}[b]{0.45\textwidth}
    \includegraphics[scale=0.45,width=2.8in]{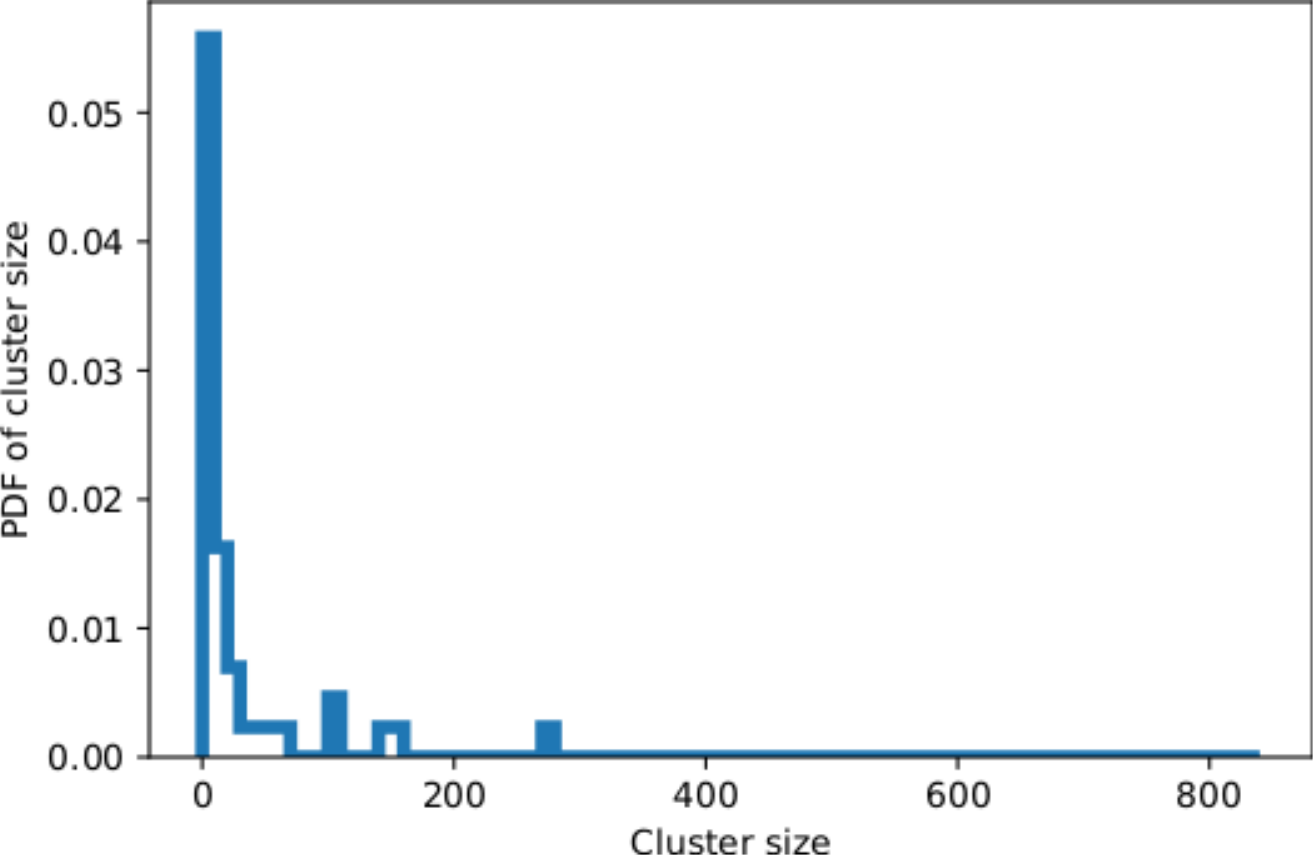}
      \caption{$\sigma=5\times10^{-4}$} 
    \end{subfigure} 
%\end{center}
    \caption{Particle cluster size (i.e., the total number of particles in a cluster) distribution in a time span from $10^6$ to $4 \times 10^6 \Delta t$. Clusters form due to the presence of capillary bridges between the suspended particles in a shear flow (a: no capillary bridges force; b,c,d: capillary force with different surface tension parameter $\sigma$).}
\label{fig9}
\end{figure*}

Model C predicts the capillary force with high accuracy, and is also easy to be implemented in a DEM code. To show the possibilities opened by coarse-graining the capillary interaction using this simple model potential, we run several LBM/DEM simulations of a suspension of particles in a shear flow, and study the influence of the presence and strength of the capillary bridge on the structural properties of the wet granular material. 

Our starting configurations are generated by placing 2000 particles (with radius $R_p=3\Delta x$) randomly in a cubic simulation box (side length of 128$\Delta x$). This corresponds to a packing fraction of 10$\%$. We assume that the secondary fluid is, effectively, uniformly distributed among all the particles with a constant fraction of 5$\times 10^{-6}$. 
 
Capillary bridge forces are introduced using model C and, consequently, the dynamics of the secondary fluid is not resolved in these simulations. The contact angle parameter is set to 123 degrees. We obtain different degrees of capillary bridge strength by changing the surface tension coefficient. The shear flow is imposed using Lees-Edwards boundary conditions~\cite{lees1972computer,HVC04} in the $x$-direction, generating a spatially homogeneous, linear shear flow. The usual periodic boundary conditions with no imposed velocity are applied to the remaining directions.

In Fig~\ref{fig8}, we present instantaneous snapshots of the system, showing the formation of particle clusters due to capillary bridges. We colour the particles based on a simple cutoff clustering algorithm, as implemented in the Pytim software package~\cite{sega2018pytim}. Two particles are assigned to the same cluster if their distance is less than the cutoff $\delta_r=6.5\Delta x$ (the radius of particles being 3.0$\Delta x$). Without capillary bridge forces (Fig~\ref{fig8}a), particles are homogeneously distributed in the system. With weak capillary bridge forces (Fig~\ref{fig8}b), only a limited tendency to cluster formation is observed. The aggregating force is balanced by the presence of shear, that tends to disrupt the clusters.  With stronger capillary forces (Figs~\ref{fig8}c and ~\ref{fig8}d), the shear forces are not anymore able to prevent the formation of clusters, and increasingly large structures are found.

To provide a more quantitative picture, in Fig~\ref{fig9} we report the histograms of the cluster size distribution, sampled over the time span from $10^6$ to $4 \times 10^6 \Delta t$, after having reached stationary conditions. Without capillary force (Fig~\ref{fig9}a), around 80$\%$ of the particles do not belong to any cluster (cluster size equal one), and the largest cluster observed is composed of 8 particles. With increasing capillary force (Figs~\ref{fig9}b-d), an increasing amount of particles is involved in larger clusters and eventually, particles tend to agglomerate in a single cluster (Fig~\ref{fig9}d, the average size of the largest cluster containing 850 particles). In this case, the particles not belonging to any cluster are just a small fraction (5-6\%) of the total number of clusters.

\section*{Conclusion}

We performed lattice Boltzmann numerical simulations of the static capillary bridge between equally and unequally-sized spherical particles at different wetting conditions and liquid bridge volumes. We reviewed some of the most popular analytical and semi-analytical models that predict the capillary bridge force between particles and compared them to the results of the numerical simulations, to explore their validity and suitability for modeling capillary bridges forces.

For pairs of equally-sized spherical particles, sufficiently small contact angles and
liquid-to-solid volume ratios, three models are found to be in good qualitative and quantitative agreement with the numerical simulations. 
In case of unequal spherical particles, the model of Willett and coworkers~\cite{willett2000capillary} (Model C), as well that of Chen and coworkers~\cite{chen2011liquid} (Model F) turned out to describe reasonably well the capillary bridge force, also at moderately high contact angles and volume fractions. In this sense, both models are good candidates to implement capillary forces in discrete element method simulations. 

As an example of a possible application, we showed the capability of model C to study the influence of the capillary bridge force on the agglomeration of particles in a shear flow. Up to moderate capillary force strengths, the imposed shear flow is able to disrupt the forming clusters, but with stronger interactions one observes the transition to an almost complete agglomeration. 

This work lays the foundation of future investigations on the formation and rheology of large-scale systems of capillary-bridge agglomerates, thanks to the reduced computational requirement of the coarse-grained model.
Possible future extensions of the current work include the modelling of asymmetric bridges, which form at high contact angles, and the inclusion of liquid transport across the bridges.

\section*{Acknowledgement}

The authors thank B.~Nun, T.~Plankenb\"uhler and J.~Karl for fruitful discussions. Financial support by the German Research Foundation (DFG) within the project HA\,4382/7-1 as well as the computing
time granted by the J\"ulich Supercomputing
Centre (JSC) are highly acknowledged.

\section*{Appendix}

The definitions of the coefficients $f_1, f_2, f_3$ and $f_4$ in Table~\ref{table2} for model C are in Table~\ref{table8}.

\begin{table*}
\begin{center}
\setlength\arrayrulewidth{1.0pt}
    \begin{tabular}{|c| c|}
    \hline
    Coefficients & Expressions  \\ \hline
    $f_1$
    &$(-0.44507 + 0.050832 \theta - 1.1466\theta^2 )$\\
    &$- (0.1119 + 0.000411\theta + 0.1490\theta^2 ) \ln \Bar{V} $\\
    &$-(0.012101 + 0.0036456\theta + 0.01255\theta^2 )(\ln \Bar{V})^2$ \\
    &$- (0.0005 + 0.0003505\theta + 0.00029076\theta^2 )(\ln \Bar{V})^3$\\
    \hline
    $f_2$
    & $(1.9222 - 0.57473\theta- 1.2918\theta^2 ) $\\
    & $- (0.0668 + 0.1201\theta + 0.22574\theta^2 )\ln \Bar{V} $\\
    & $- (0.0013375 + 0.0068988\theta + 0.01137\theta^2 )(\ln \Bar{V})^2$\\
     \hline
    $f_3$
    & $ 1.268 - 0.01396\theta - 0.23566\theta^2$ \\
    &$+(0.198 + 0.092\theta - 0.06418\theta^2 )\ln \Bar{V}$ \\
    &$+(0.02232 + 0.02238\theta - 0.009853\theta^2 )(\ln \Bar{V})^2$ \\
    &$+(0.0008585 + 0.001318\theta- 0.00053\theta^2 )(\ln \Bar{V})^3$\\
     \hline
    $f_4$
    &$-0.010703 + 0.073776\theta- 0.34742\theta^2 $\\
    &$+(0.03345 + 0.04543\theta - 0.09056\theta^2 )\ln \Bar{V} $\\
    &$+(0.0018574 + 0.004456\theta - 0.006257\theta^2 )(\ln \Bar{V})^2$\\
    \hline
    \end{tabular}
\caption{Coefficients $f_1, f_2, f_3$ and $f_4$ in Table \ref{table2} for Model C ($\Bar{V} = \frac{4\pi}{3}V^*$)}
\label{table8}
\end{center}
\end{table*}

\bibliography{citation.bib}

\clearpage
\onecolumngrid
\begin{center}
{\bf 
\Large   Supplementary Material\\[2em]
}
\end{center}
\clearpage

\begin{figure*}[tbp]
 \captionsetup{justification=raggedright}
  \begin{subfigure}[b]{0.45\textwidth}
    \includegraphics[scale=0.45,width=2.0in]{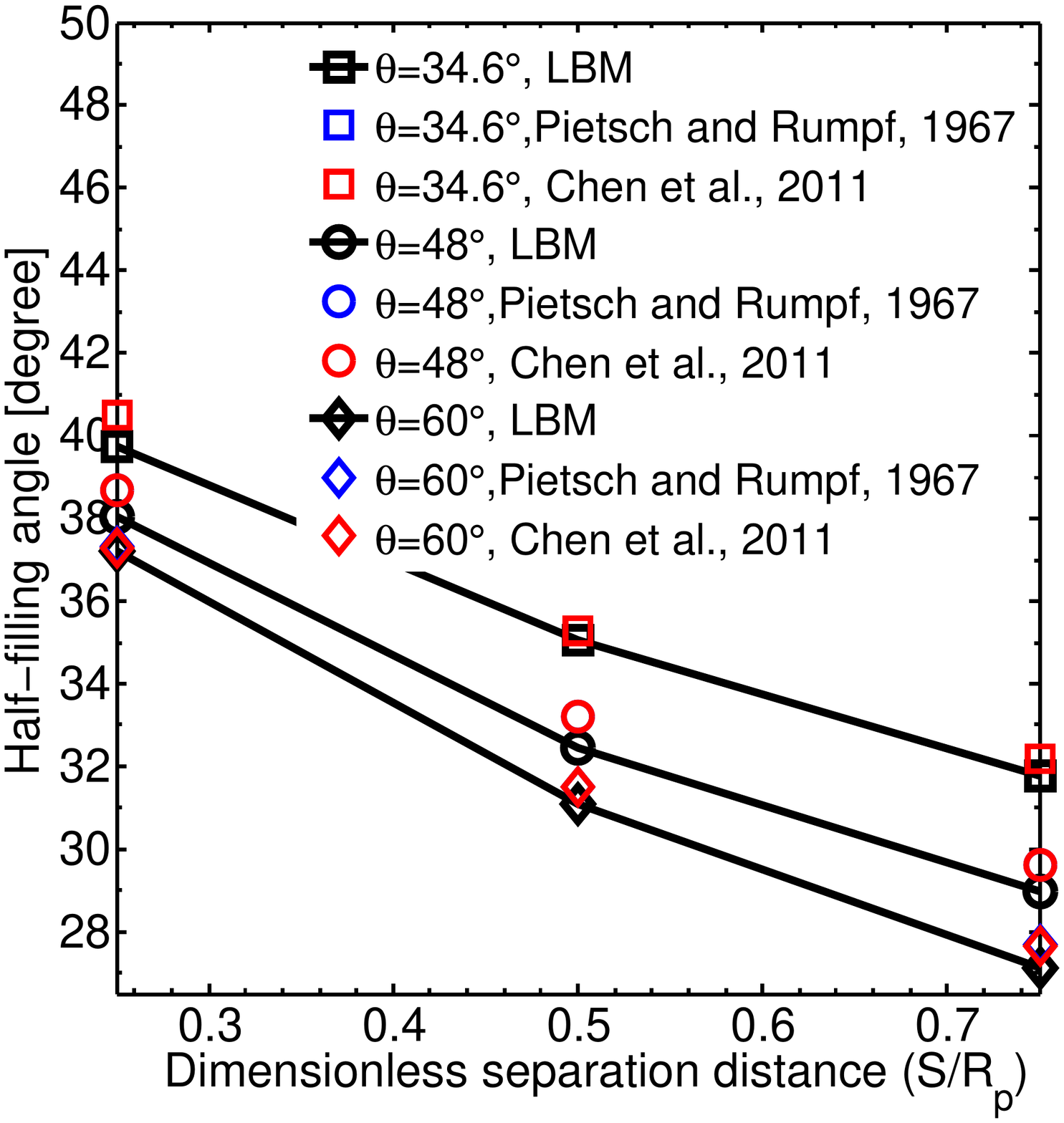}
       \caption{Different contact angles ($V^*=0.133$)} 
  \end{subfigure}
  \hspace{3mm}
   \begin{subfigure}[b]{0.45\textwidth}
    \includegraphics[scale=0.45,width=2.0in]{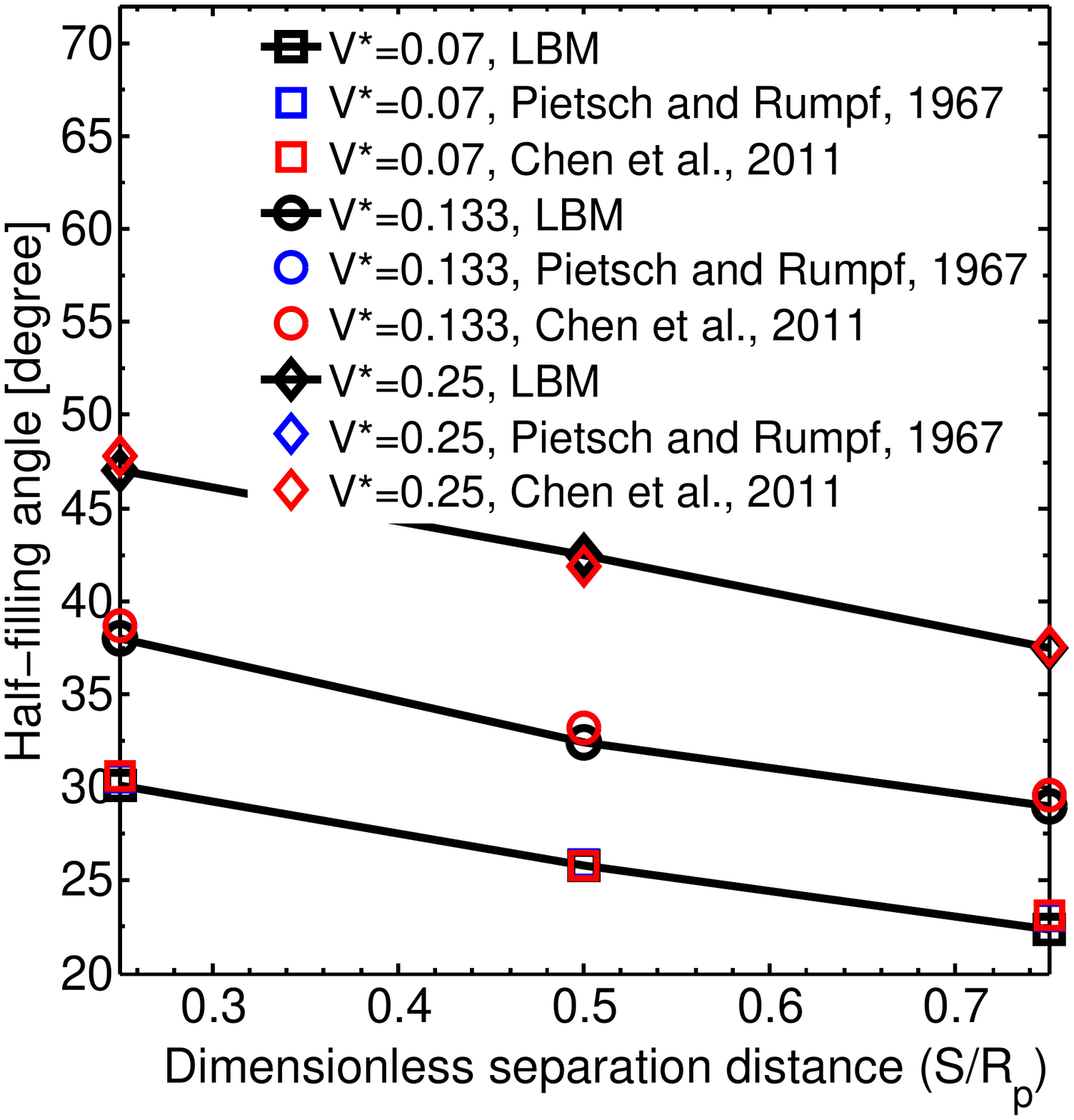}
      \caption{Different bridge volumes ($\theta=48^\circ$)} 
    \end{subfigure} 
    \caption{Half-filling angle as a function of the separation
    distance under different contact angles and liquid bridge volumes
    between two equal spheres.}
\label{fig10}
\end{figure*}

Additional comparisons of lattice Boltzmann results with theoretical models are shown in Fig.~\ref{fig10} and Tables \ref{table4}--\ref{table9}.

Figs.~\ref{fig10}(a,b) display the half-filling angle as a function
of the separation distance under different contact angles and liquid
volumes. The predicted bridge filling angles from lattice Boltzmann simulations
and the theoretical models from Pietsch and
Rumpf~\cite{pietsch1967haftkraft} and Chen and coworkers~\cite{chen2011liquid}
match well. We make comparisons between lattice Boltzmann simulations and theoretical model
from Pietsch and Rumpf~\cite{pietsch1967haftkraft} for the prediction
of the liquid bridge volume in Table~\ref{table4}. Note that the
model from Megias-Alguacil and Gauckler~\cite{megias2009capillary}
is fully equivalent to Pietsch and Rumpf's model for the calculation
of the liquid bridge volume.

\begin{table*}
	\begin{center}
\setlength\arrayrulewidth{1.0pt}
    \begin{tabular}{|l|l l l l |l l l l|}
    \hline
    Cases & \multicolumn{4}{c|}{Bridge volume ($V_{T}^*$)} & \multicolumn{4}{c|}{Error ($\frac{V_{LB}^*-V_{T}^*}{V_{T}^*}$) \%} \\ \hline
    \multicolumn{9}{|c|}{Large $V_l$} \\ \hline
    & $S^*$=0.286 &$S^*$=0.572 &$S^*$=0.858 &$S^*$=1.144 &$S^*$=0.286 &$S^*$=0.572 &$S^*$=0.858 &$S^*$=1.144  \\ 
    $\theta=34.6^\circ$ &0.418 &0.534 &0.627 &0.622 &-4.4 &-6.8 &-7.9 &2.44  \\ 
    $\theta=48^\circ$   &0.382 &0.47 &0.557 &0.62   &2.66 &3.13 &3.59 &2.8  \\ 
    $\theta=60^\circ$   &0.366 &0.485 &0.583 &0.657 &5.11 &-1.1 &-2.1 &-3.5 \\ 
    $\theta=74^\circ$   &0.373 &0.503 &0.531 &0.603 &-0.2 &-5.8 &6.37 &4.22 \\\hline
    Cases & \multicolumn{4}{c|}{Bridge volume ($V_{T}^*$)} & \multicolumn{4}{c|}{Error ($\frac{V_{LB}^*-V_{T}^*}{V_{T}^*}$) \%} \\ \hline
  \multicolumn{9}{|c|}{Small $V_l$} \\ \hline
    &$S^*$=0.286 &$S^*$=0.428 &$S^*$=0.572 &$S^*$=0.858 &$S^*$=0.286 &$S^*$=0.428 &$S^*$=0.572 &$S^*$=0.858 \\
    $\theta=34.6^\circ$ &0.123 &0.098 &0.103 &0.151 &8.6  &5.6 &4.7 &3.8 \\
    $\theta=48^\circ$   &0.107 &0.101 &0.093 &0.145 &-2.6 &4.5 &6.9 &8.0 \\ 
    $\theta=60^\circ$   &0.097 &0.072 &0.075 &0.127 &7.0  &2.9 &0.7 &2.4 \\ 
    $\theta=74^\circ$   &0.096 &0.055 &0.055 &0.105 &-5.7 &9.0 &-0.9 &5.5  \\
    \hline
    \end{tabular}
\caption{Comparison of the dimensionless bridge volume ($V^*=V_l/V_p$, $V_T^*$ from Pietsch and Rumpf, 1967, $V_{LB}^*$ from  lattice Boltzmann simulation) for equal spheres from theoretical models and Lattice Boltzmann simulations}
\label{table4}
\end{center}
\end{table*}
Tables \ref{table5} and \ref{table6} show the capillary
force obtained from lattice Boltzmann simulations and the theoretical models for equal spheres. Within the range of parameters tested, the solutions by lattice Boltzmann and the
theoretical models agree qualitatively, apart from bigger deviations
at large contact angles (e.g., at $\theta = 60^\circ, 74^\circ$ ). 

\begin{table*}
\begin{center}
\setlength\arrayrulewidth{1.0pt}
    \begin{tabular}{|l|l l l l|l l l l|l l l l|}
    \hline
    \multirow{2}{*}{Cases} & \multicolumn{4}{c|}{$F^*$ from LB} & %
    \multicolumn{4}{c|}{$\frac{F_{LB}^*-F_{A}^*}{F_{A}^*}$(\%, Model A)} & \multicolumn{4}{c|}{ $\frac{F_{LB}^*-F_{B}^*}{F_{B}^*}$ (\%, Model B)}\\
\cline{2-13}
 & \multicolumn{4}{c|}{Separation $S^*$} & \multicolumn{4}{c|}{Separation $S^*$} & \multicolumn{4}{c|}{Separation $S^*$}\\
\cline{2-13}
 &0.286 &0.572 &0.858 &1.144 &0.286 &0.572 &0.858 &1.144& 0.286 &0.572 &0.858 &1.144 \\ 
    \hline
    $\theta=19.2^\circ$&3.03&2.87 &2.7 &2.52 &-0.75 &-3.8 &-6.9 &-5.2 &-0.6&-3.4 &-6.4 &-5.3\\ 
    $\theta=34.6^\circ$&2.447&2.404 &2.307 &2.22 &-3.0 &-5.2&-6.9 &-6.6 &-2.9&-5.2 &-6&-6.8\\ 
    $\theta=48^\circ$&1.754 &1.849 &2.026 &2.018 &-4.42&-11.5 &-7.9 &-9.5 &-4.0 &-11.1 &-7.6 &-9.3 \\ 
    $\theta=60^\circ$&0.833&1.529 &1.708 &1.726 &-22.1&-14.3 &-14.0 &-12.1 &-21.2&-13 &-12.5 &-10.5 \\ 
 \hline
    \multirow{3}{*}{Cases} & \multicolumn{4}{c|}{$\frac{F_{LB}^*-F_{C}^*}{F_{C}^*}$(\%, Model C)} & %
    \multicolumn{4}{c|}{$\frac{F_{LB}^*-F_{D}^*}{F_{D}^*}$(\%, Model D)} & \multicolumn{4}{c|}{$\frac{F_{LB}^*-F_{E}^*}{F_{E}^*}$ (\%, Model E)}\\
\cline{2-13}
 & \multicolumn{4}{c|}{Separation $S^*$} & \multicolumn{4}{c|}{Separation $S^*$} & \multicolumn{4}{c|}{Separation $S^*$}\\
\cline{2-13}
 &0.286 &0.572 &0.858 &1.144 &0.286 &0.572 &0.858 &1.144& 0.286 &0.572 &0.858 &1.144 \\ 
    \hline
    $\theta=19.2^\circ$&-2.4 &2.74&6.48 &9.24 &-27.5 &-11.3 &3.6 &21.8 &-27.4 &-24.4 &-22.7 &-22.2 \\
    $\theta=34.6^\circ$&1.6 &11.5 &18.1 &17.6 &-36.5 &-15.9  &6.7 &29.5 &-23.8 &-14.3 &-7.7 &-3.9 \\ 
    $\theta=48^\circ$&6.2 &19.9 &38.0 &29.7 &-42.9 &-20.2 &10.4 &38.9 &-22.5 &-6.73 &13.32 &23.1 \\ 
    $\theta=60^\circ$&-20.6 &29.5 &56.9 &42.6 &-63.6 &-37.9 &11.8 &57.2 &-45.9 &-2.9 &29 &57.6 \\ 
    \hline
    \end{tabular}
\caption{Comparison of the dimensionless capillary forces ($F^*=F/(\sigma R_p)$; $F_{A}^*$: $F^*$ from model A; $F_{B}^*$: $F^*$ from model B; $F_{C}^*$: $F^*$ from model C; $F_{D}^*$: $F^*$ from model D; $F_{E}^*$: $F^*$ from model E) for equal spheres from theoretical models and Lattice Boltzmann simulations with relatively large liquid  volume ($V^*=0.4$ for $S^*=0.286$, $V^*=0.5$ for $S^*=0.572$, $V^*=0.59$ for $S^*=0.858$, $V^*=0.65$ for $S^*=1.144$)}
\label{table5}
\end{center}
\end{table*}
\begin{table*}
\begin{center}
\setlength\arrayrulewidth{1.0pt}
    \begin{tabular}{|l|l l l |l l l |l l l| }
    \hline
    \multirow{3}{*}{Cases} & \multicolumn{3}{c|}{$F^*$ from LB} & %
    \multicolumn{3}{c|}{$\frac{F_{LB}^*-F_{A}^*}{F_{A}^*}$(Model A)} & \multicolumn{3}{c|}{$\frac{F_{LB}^*-F_{B}^*}{F_{B}^*}$ (Model B)}\\
\cline{2-10}
 & \multicolumn{3}{c|}{Separation $S^*$} & \multicolumn{3}{c|}{Separation $S^*$} & \multicolumn{3}{c|}{Separation $S^*$}\\
\cline{2-10}
 &0.286 &0.572 &0.858 &0.286 &0.572 &0.858 & 0.286 &0.572 &0.858 \\ 
    \hline
    $\theta=34.6^\circ$&2.527 &1.852 &1.64 &-1.73\%&-0.8\% &-2.73\% &-0.68\% &-4.21\% &-5.7\%\\ 
    $\theta=48^\circ$ &2 & 1.583 &1.516 &-5.5\% &-3.11\% &-9.06\% &-5.43\% &-1.66\% &-7.74\%\\ 
    $\theta=60^\circ$ &1.324 &1.216 &1.242 &-15.85\% &-9.55\% &-2.88\% &-15.83\% &-9.31\% &-2.36\%\\ 
    $\theta=74^\circ$ &0.747 &0.859 &1.016 &-13.81\% &-15.32\% &-2.86\% &-13.41\% &-15.25\% &-2.69\%\\\hline
    \multirow{3}{*}{Cases} & \multicolumn{3}{c|}{$\frac{F_{LB}^*-F_{C}^*}{F_{C}^*}$(Model C)} & %
    \multicolumn{3}{c|}{$\frac{F_{LB}^*-F_{D}^*}{F_{D}^*}$(Model D)} & \multicolumn{3}{c|}{ $\frac{F_{LB}^*-F_{E}^*}{F_{E}^*}$ (Model E)}\\
\cline{2-10}
 & \multicolumn{3}{c|}{Separation $S^*$} & \multicolumn{3}{c|}{Separation $S^*$} & \multicolumn{3}{c|}{Separation $S^*$}\\
\cline{2-10}
 &0.286 &0.572 &0.858 &0.286 &0.572 &0.858 & 0.286 &0.572 &0.858 \\ 
    \hline
    $\theta=34.6^\circ$&-0.66\% &-3.81\% &-0.17\% &-17.4\% &25.94\% &15.87\% &-25.23\% &-32.37\% &-30.12\%\\ 
    $\theta=48^\circ$  &2.37\% &-2.41\% &2.5 \% &-19.6\% &38.33\% &53.67\% &-25.94\% &-28.13\% &-17.6\%\\ 
    $\theta=60^\circ$  &-7.86\% &-8.51\% &-3.91\% &-20.58\% &67.11\% &50.3\% &-38.58\% &-31.48\% &-11.53\%\\ 
    $\theta=74^\circ$  &-19.24\% &-22.47 \%&-9.86\% &-14.76\% &66.11\% &60.57\% &-53.93\% &-40.8\% &1.26\% \\
    \hline
    \end{tabular}
\caption{Comparison of the dimensionless capillary forces ($F^*=F/(\sigma R_p)$; $F_{A}^*$: $F^*$ from model A; $F_{B}^*$: $F^*$ from model B; $F_{C}^*$: $F^*$ from model C; $F_{D}^*$: $F^*$ from model D; $F_{E}^*$: $F^*$ from model E) for equal spheres from theoretical models and Lattice Boltzmann simulations with relatively small liquid volume ($V^*=0.09$ for $S^*=0.286$, $V^*=0.12$ for $S^*=0.572$, $V^*=0.15$ for $S^*=0.858$)}
\label{table6}
\end{center}
\end{table*}
We compare the liquid bridge volume as obtained from lattice Boltzmann simulations and theoretical
models by Chen and coworkers~\cite{chen2011liquid} and Sun and
Sakai~\cite{sun2018liquid} in Table~\ref{table7}.
\begin{table*}
\begin{center}
\setlength\arrayrulewidth{1.0pt}
    \begin{tabular}{|l |l l l|l l l|l l l| }
    \hline
    \multirow{3}{*}{Cases} & \multicolumn{3}{c|}{$V^*_{L}$ from LB} & %
    \multicolumn{3}{c|}{$\frac{V_{LB}^*-V_{Chen}^*}{V_{Chen}^*}$(Chen et al.)} & \multicolumn{3}{c|}{$\frac{V_{LB}^*-V_{S}^*}{V_{S}^*}$ (Sun and Sakai)}\\
\cline{2-10}
 & \multicolumn{3}{c|}{Contact angle $\theta$} & \multicolumn{3}{c|}{Contact angle $\theta$} & \multicolumn{3}{c|}{Contact angle $\theta$}\\
\cline{2-10}
 &34.6$^\circ$ &48$^\circ$ &60$^\circ$ &34.6$^\circ$ &48$^\circ$ &60$^\circ$ &34.6$^\circ$ &48$^\circ$ &60$^\circ$ \\
 \hline
    $S^*$=0.194 &0.091 &0.085 &0.074 &2.11\%  &7.82\%  &-2.63\% &4.9\%   &8.29\%  &1.14\% \\ 
    $S^*$=0.364 &0.098 &0.089 &0.081 &3.56\%  &5.20\%  &8.43\%  &-4.20\% &-13.9\% &-5.2\%  \\ 
    $S^*$=0.486 &0.112 &0.108 &0.099 &-2.56\% &4.0\%   &-8.29\% &6.03\%  &--1.2\% &9.37\%  \\
    $S^*$=0.68  &0.128 &0.127 &0.118 &-6.85\% &-4.32\% &-0.47\% &10.69\% &4.53\%  &-5.1\%\\
    $S^*$=0.85  &0.130 &0.130 &0.121 &-9.81\% &-8.64\% &-9.23\% &12.22\% &0\% &5.63\%\\ \hline
    \end{tabular}
\caption{Comparison of the dimensionless bridge volume ($V^*=V_l/V_p$; $V_{Chen}^*$: $V^*$ from Chen et al.(\cite{chen2011liquid}); $V_{S}^*$: $V^*$ from Sun and Sakai(\cite{sun2018liquid})) for unequal spheres from theoretical models and Lattice Boltzmann simulations.}
\label{table7}
\end{center}
\end{table*}
Table~\ref{table8} depicts the dimensionless capillary forces at
different contact angles and separation distances. We observe good agreement between lattice Boltzmann simulations and the theoretical models with small relative errors for the predictions of the bridge volume and capillary force for different cases.

\begin{table*}
\begin{center}
\setlength\arrayrulewidth{1.0pt}
    \begin{tabular}{|l|l l l|l l l|l l l|}
    \hline
    \multirow{3}{*}{Cases} & \multicolumn{3}{c|}{$F^*$ from LB} & %
    \multicolumn{3}{c|}{$\frac{F_{LB}^*-F_{C}^*}{F_{C}^*}$( Model C)} & \multicolumn{3}{c|}{$\frac{F_{LB}^*-F_{Chen}^*}{F_{Chen}^*}$ (Chen et al.)}\\
\cline{2-10}
 & \multicolumn{3}{c|}{Contact angle $\theta$} & \multicolumn{3}{c|}{Contact angle $\theta$} & \multicolumn{3}{c|}{Contact angle $\theta$}\\
\cline{2-10}
 &34.6$^\circ$ &48$^\circ$ &60$^\circ$ &34.6$^\circ$ &48$^\circ$ &60$^\circ$ &34.6$^\circ$ &48$^\circ$ &60$^\circ$ \\
 \hline
    $S^*$=0.194  &2.631 &1.981 &1.371 &-3.4\%  &-7.55\%  &-14.82\% &-3.8\%  &-5.65\% &-10.9\% \\ 
    $S^*$=0.364 &2.205 &1.789 &1.36  &-6.52\% &-11.35 &-7.86\%  &-3.92\% &-2.77\% &-4.97\%  \\ 
    $S^*$=0.486 &1.99  &1.708 &1.367 &-3.06\% &-7.34  &-4.41\%  &-4.05\% &-0.56\% &-0.38\%  \\
    $S^*$=0.68 &1.755 &1.569 &1.345 &2.8\%  &-4.05  &-6.11\%  &-1.94\% &0.7\%   &2.1\%\\
    $S^*$=0.85 &1.494 &1.404 &1.202 &5.38\% &-2.74  &3.6\%    &-2.86\% &-1.31\% &-0.623\%\\ \hline
    \end{tabular}
\caption{Comparison of the dimensionless capillary forces ($F^*=F/(\sigma R_p)$; $F_{C}^*$: $F^*$ from model C\cite{willett2000capillary}; $F_{Chen}^*$: $F^*$ from Chen et al.\cite{chen2011liquid} and Sun and Sakai~\cite{sun2018liquid}) for unequal spheres from theoretical models and Lattice Boltzmann simulations.}
\label{table9}
\end{center}
\end{table*}

\end{document}